\documentclass[prc,twocolumn,amsmath,amssymb,superscriptaddress,noffotinbib,preprintnumbers]{revtex4-1}
\usepackage{url,color,colordvi,epsfig,pstricks}
\usepackage[colorlinks=true, citecolor=blue, filecolor=blue, linkcolor=blue, urlcolor=blue, pdftex]{hyperref}
\usepackage{graphicx,bm,dcolumn}
\usepackage{hyperref,slashed,lineno}
\usepackage{physics}
\usepackage{inputenc}

\hypersetup{
    colorlinks=true, 
    linktoc=all, 
    linkcolor=blue, 
    citecolor=blue}

\newcommand{\cgc}[6]{ \left( #1 \: #2 \; #3 \: #4 \right| #5 \: #6 \left.\right) }

\newcommand{\diff}{\mathrm{d}}						 								

\begin{document}

\title{Assessing the theory-data tension in neutrino-induced charged pion production: the effect of final-state nucleon distortion}
\author{A.~Nikolakopoulos}\email{anikolak@fnal.gov}
\affiliation{Theoretical Physics Department, Fermilab, Batavia IL 60510, USA}
\author{R.~Gonz\'alez-Jim\'enez}
\affiliation{Grupo de F\'isica Nuclear, Departamento de Estructura de la Materia, F\'isica T\'ermica y Electr\'onica, Facultad de Ciencias F\'isicas, Universidad Complutense de Madrid and IPARCOS, CEI Moncloa, Madrid 28040, Spain}
\author{N.~Jachowicz}
\affiliation{Department of Physics and Astronomy, Ghent University, B-9000 Gent, Belgium}
\author{J.~M.~Ud\'ias}
\affiliation{Grupo de F\'isica Nuclear, Departamento de Estructura de la Materia, F\'isica T\'ermica y Electr\'onica, Facultad de Ciencias F\'isicas, Universidad Complutense de Madrid and IPARCOS, CEI Moncloa, Madrid 28040, Spain}


\begin{abstract}
\begin{description}
\item[Background] 
Pion production on nuclei constitutes a significant part of the total cross section in experiments involving few-GeV neutrinos.
Combined analyses of data on deuterium and heavier nuclei points to tensions between the bubble chamber data and the data of the MINER$\nu$A experiment, which are often ascribed to unspecified nuclear effects.
\item[Purpose] In experimental analysis use is made of approximate treatments of nuclear dynamics, usually in a Fermi gas approach with classical treatments of the reaction mechanism, and fits are often performed by simply rescaling cross sections. 
To understand the origin of these tensions, check the validity of approximations, and to further advance the description of neutrino pion production on nuclei, a microscopic quantum mechanical framework is needed to compute nuclear matrix elements.
\item[Method] We use the local approximation to the relativistic distorted wave impulse approximation (RDWIA) to calculate nuclear matrix elements.
We include the distortion of wave functions of the final-state nucleon in a real energy-dependent potential. We compare results with and without distortion.
To perform this comparison under conditions relevant to neutrino experiments, we compute cross sections for the MINER$\nu$A and T2K charged pion production datasets.
\item[Results] The inclusion of nucleon distortion leads to a reduction of the cross section up to 10\%, but to no significant change in shape of the flux-averaged cross sections. 
Results with and without distortion compare favorably to experimental data, with the exception of the low-$Q^2$ MINER$\nu$A $\pi^+$ data.
We point out that hydrogen target data from BEBC is also overpredicted at low-$Q^2$, and the data-model discrepancy is similar in shape and magnitude as what is found in comparison to MINER$\nu$A data.
\item[Conclusions]
Including nucleon distortion alone cannot explain the overprediction of low-$Q^2$ cross sections measured by MINER$\nu$A. 
The similar overprediction of BEBC data on hydrogen means that it is impossible to ascribe this discrepancy solely to a nuclear effect. 
Axial form factors might not be constrained in a satisfactory way by the ANL/BNL data alone.
Axial couplings and their $Q^2$ dependence should ideally be derived from more precise data on hydrogen and deuterium.
Nuclear matrix elements should be tested with e.g. electron scattering data for which nucleon level physics is better constrained.
\end{description}
\end{abstract}

\maketitle


\section{Introduction}
In neutrino experiments such as DUNE and NO$\nu$A inelastic interactions with the nucleon constitutes a large part of the total event rate~\cite{NOvA:2020Tune, NOVA:DDIF2021}. 
Experiments such as T2K, MiniBooNE, MicroBooNE and the short-baseline program at Fermilab are more sensitive to quasi-elastic scattering and meson-exchange currents.
However these experiments are still sensitive to inelastic processes, mostly single pion production (SPP) in the $\Delta$ region~\cite{MiniBooNECCpi010,T2Kinc13,MicroBooNE:2022lvx, MicroBooNE:2019nio}.
Experiments often adopt a signal definition in which events where a pion is present are rejected, but this procedure does not not fully remove inelastic contributions~\cite{Lalakulich12b,Dolan:2021rdd,CLAS:2021neh}.
The inelastic contribution to a $0\pi$ signal is often labeled `pion-absorption', but may also consist of pions below detector thresholds, or non-pionic decays of resonances. Comparisons to electron scattering data show that this region is problematic in the commonly used GENIE event generator, due to double counting of resonance and DIS descriptions~\cite{CLAS:2021neh, Ankowski:2020qbe}.

Over the past couple of years measurements of cross sections for neutrino-induced pion production on nuclei have been performed. 
Notably the MINER$\nu$A experiment has reported cross sections for both neutrino and antineutrino production of both neutral and charged pions on carbon~\cite{MINERvACCpi15,MINERvACCpi015, MINERvACCpi16, MINERvAdiff16, MINERvAnuCC17pi0, Minerva:CCnuanu2017,MINERvA:pimin2019}.
Analysis of the different MINER$\nu$A datasets in Ref.~\cite{ANLBNLMinerva} highlights that there are tensions between the different datasets and with the ANL/BNL hydrogen and deuterium bubble chamber data.
A notable \emph{ad-hoc} modification introduced to improve model-data agreement is a large suppression of the cross section at low four-momentum transfer $Q^2$, although the amount of suppression needed varies with the dataset.
A similar modification was used by the NO$\nu$A collaboration in fitting their measurements of the inclusive cross section~\cite{NOvA:2020Tune}.
These modifications are applied to the interactions on nuclei only, motivated by the analogy to quasielastic interactions.
In that case it is indeed established that inclusion of Pauli blocking, the distortion of the final state, long and short-range correlations yield smaller cross sections at low-$Q^2$ compared to equivalent calculations that omit these effects~\cite{Gonzalez-Jimenez19, Nieves:2017lij}.
It is important to actually compute the effect of these mechanisms, based on established microscopic approaches, which can be validated with different interaction mechanisms.
Empirical fits based on a limited amount of data, the interpretation of which might be highly model dependent, is unsatisfactory and can bias analyses.

This becomes especially important as the amplitudes for neutrino SPP on nucleons, that serve as input to nuclear models, are poorly constrained~\cite{NuSTEC:2020nsl, NuSTECWP, SajjadAthar:2022pjt}.
While theoretical approaches for electroweak amplitudes may differ in sophistication, models for electroweak SPP amplitudes that span the delta region and beyond rely on the analysis of precise data of differential cross sections, which is available only for electromagnetic pion production.
For the neutrino-induced processes such high-quality data does not exist, and one has to rely on total and flux-averaged cross sections obtained in deuterium and hydrogen bubble chambers~\cite{CC-ANL82_long, CC-BNL86, ALLASIA1990}.
As such it is hard, if not impossible, to identify whether model-data discrepancies in current flux-averaged neutrino cross sections are due to any specific mechanism, be it nucleus or nucleon specific.

The relativistic distorted wave impulse approximation (RDWIA) provides the framework of choice to compute nuclear effects in direct pion production reactions, where a pion is produced through a single nucleon operator.
The RDWIA and non-relativistic DWIA, have proven to be successful tools for interpreting and describing nucleon knockout in electromagnetic interactions~\cite{Atkinson:2018nvp, Kelly05, Udias93} and photon and electron-induced pion production~\cite{LAGET197281, PhysRevC.48.816, VanderhaeghenPhD}. The relativistic formulation of the RDWIA has the advantage that the full Dirac structure of the single nucleon operators is retained.
One does not need a non-relativistic reduction, which is usually obtained by cutting off the expansion of a current between free-nucleon spinors in orders of $p/M_N$ \cite{BLOMQVIST1977405, PhysRev.78.29}. 
Additionally, relativistic models for the nucleus based on density functional theory such as the relativistic mean field (RMF)~\cite{Serot86, Sharma93}, although they are phenomenological, provide an excellent description of many nuclear phenomena 
 with relatively few free parameters~\cite{RINGRMF, Serot86, BenderReinhard:RevModPhys}.

Matrix elements for SPP in the RDWIA require as input a single-nucleon operator, wavefunctions for the initial and final-state nucleon and pion.
We currently cannot provide all these ingredients, notably a potential for the distorted pion wavefunctions which is suitable for the experimental signature in neutrino experiments is not available. 
In this work we  isolate and tackle specifically the effect of distortion of the final-state nucleon. The outgoing nucleon wavefunctions are obtained with the energy-dependent RMF (EDRMF) potential introduced in Ref.~\cite{Gonzalez-Jimenez19}. The main appeal of this treatment is that initial and final state potentials are identical for low nucleon energies, this leads naturally to Pauli blocking~\cite{Nikolakopoulos19}, and the conservation of the Dirac current~\cite{Nikolakopoulos:2020fti}.
This consistency combined with the energy dependence leads to an excellent description of $(e,e')$ data from small to large momentum transfers in the quasielastic region~\cite{Gonzalez-Jimenez:2019ejf}.
We use the common approximation where the asymptotic value of nucleon momentum is used in evaluating the single-nucleon operator~\cite{PhysRevC.48.816, LAGET197281}. 
We perform direct comparisons of the RDWIA with the equivalent calculation in the relativistic plane wave impulse approximation (RPWIA) in order to asses the effect and importance of nucleon distortion.

This paper is structured as follows:
In Section~\ref{secCS} we describe the RDWIA formalism, first the general expression of the cross section and kinematics in the RMF are discussed. In Sec.~\ref{sec:distortion} we start from the general form of the nuclear current in RDWIA and discuss the approximations made that lead to the local RDWIA and the RPWIA expressions. Finally in Sec.~\ref{sec:SPPnucleon} we discuss the single-nucleon operator used in our calculations, and provide a comparison of the isovector contribution with more advanced analyses of electron scattering data.
The comparison of RPWIA and RDWIA results with each other and with experimental data is shown in Sec.~\ref{sec:results}. Finally we briefly illustrate the uncertain status of the delta coupling and present our conclusions in Sec.~\ref{sec:conclusions}.

\section{Single pion production on the nucleus}{\label{secCS}}
\begin{figure}
\includegraphics[width=0.48\textwidth]{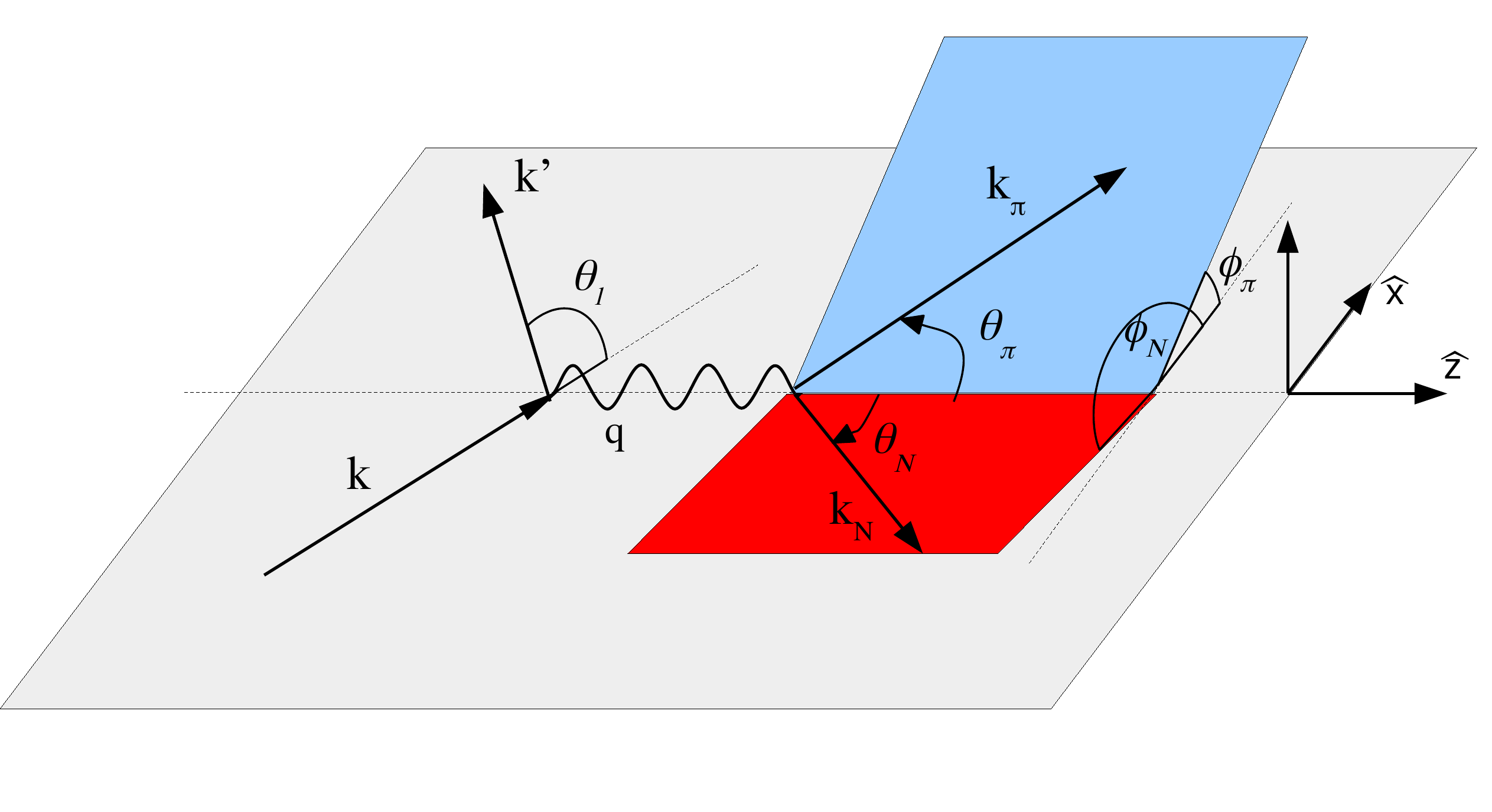}
\caption{kinematics of single pion production on the nucleus.}
\label{fig:kinematics}
\end{figure}
We consider the process of single pion production off a nucleus $A$ through a charged current interaction where a single gauge boson with four-momentum 
\begin{equation}
q^\mu = k^\mu - k^{\prime \mu},
\end{equation}
is exchanged between the lepton vertex and the hadron system.
As usual we denote the squared four momentum transfer $Q^2 = -q^2$ as positive.
We describe a "direct" reaction in which the pion is produced off a single nucleon which is excited to the continuum.
The kinematics of nucleon and pion are shown in Fig.~\ref{fig:kinematics}, the full process satisfies the conservation of the four-momentum
\begin{equation}
q^\mu + P_A^\mu = k_\pi^\mu + k_N^\mu + P_B^\mu,
\end{equation}
here, $P_A$ is the initial nucleus, and $P_B$ represents the undetected residual hadronic system.
The cross section for the charged current process can be written as
\begin{align}
\label{eq:CSpinucleus2}
&\frac{\diff^9\sigma(E)}{\diff E^\prime \diff\Omega \diff E_\pi \diff \Omega_\pi \diff E_N \diff \Omega_N} = \frac{G_F^2 \cos^2\theta_c}{2(2\pi)^8}\frac{k^\prime}{E} \frac{M_N k_\pi k_N M_B}{E_B} \nonumber \\
& \delta\left(\omega + M_A - E_\pi - E_N - E_B\right) L_{\mu\nu} H^{\mu\nu},
\end{align}
as $Q^2$ is negligible compared to the squared mass of the exchanged $W$-boson.
The lepton tensor, when the initial lepton mass is neglected, is given by
\begin{equation}
\label{eq:Lepton}
L^{\mu\nu} = k^\mu k^{\prime\nu} + k^\nu k^{\prime\mu} - g^{\mu\nu} k_\alpha k^{\prime\alpha}
- ih\epsilon^{\mu\nu\alpha\beta}k_\alpha k^{\prime}_{\beta}
\end{equation}
with $h$ the initial leptons helicity, i.e. $-1$ and $+1$ for neutrino and antineutrino reactions respectively.
The dependence on the nuclear and hadron dynamics is captured in the values $E_B$ which the residual system may take and in the hadron tensor $H^{\mu\nu}$. These are respectively described in the following subsections. 

\subsection{Kinematics}
For the following discussion we assume that the four vectors of the initial lepton $k^\mu$ and the nucleus $P_A$ are fixed.
The cross section then depends on 8 independent kinematic variables e.g. $(E^\prime, \cos\theta_l, E_\pi, \Omega_\pi, \Omega_N, M_B)$.
One sees that one produces an (unobserved) residual system with invariant mass $M_B$ and kinetic energy
\begin{equation}
\label{eq:TB}
T_B = \sqrt{M_B^2 + \left(\mathbf{p}_B\right)^2 } - M_B,
\end{equation}
with the momentum of the system (or equivalently the missing momentum $\mathbf{p}_m$ ) given by
\begin{equation}
-\mathbf{p}_m \equiv \mathbf{p}_B = \mathbf{q} - \mathbf{k}_\pi - \mathbf{k}_N.
\end{equation}
Energy conservation means that
\begin{equation}
\label{eq:Econs}
\omega + M_A = M_B + T_B + E_\pi + E_N.
\end{equation}
Given a value for $M_B$ the above equations may be solved for $E_N$, the explicit expression is given in Ref.~\cite{Gonzalez-Jimenez:2019bnz}.

In a direct knockout reaction we probe missing momenta of the order of the Fermi momentum $k_F$ such that
\begin{equation}
T_B \lesssim \frac{k_F^2}{2(A-1)M_N},
\end{equation}
when the mass of the residual system is of the order of $(A-1)M_N$. The kinetic energy becomes negligible for large nuclei.
We may hence simplify by neglecting the small recoil energy of the residual system, $T_B$ in Eq.~(\ref{eq:Econs}).

Model-dependence comes in when considering the values that $M_B$ can take.
We consider the interaction with a nucleon within the RMF shell model, for which the initial-state nucleus is described as a Slater determinant of single-particle orbitals.
The single particle states are characterized by isospin projection, a principal quantum number $n$, relativistic angular momentum $\kappa$ and the projection of total angular momentum $m_j$.
Due to spherical symmetry the $m_j$ states for fixed $n,\kappa$ are energy-degenerate.
The single-particle energy $E_{n,\kappa}$ is the energy needed to excite a nucleon from an $(n,\kappa)$ state to the continuum.
As we neglect $T_B$ we thus have
\begin{equation}
T_N = \omega - T_\pi - E_{n,\kappa}.
\end{equation}
This approach implies that the residual system is left in an internally excited state with invariant mass $M_B = E_{n,\kappa} + M_A - M_N$. 
Neglecting the nuclear recoil, summing over the possible $(n,\kappa)$ states and integrating over the outgoing nucleon energy the cross section becomes
\begin{equation}
\label{eq:CSpinucleus}
\frac{\diff^8\sigma(E)}{\diff E^\prime \diff\Omega \diff E_\pi \diff \Omega_\pi \diff \Omega_N} = \frac{G_F^2\cos^2\theta_c}{2\left(2\pi\right)^8}\frac{k^\prime}{E} M_N k_\pi L_{\mu\nu} \sum_{\kappa} k_{N,\kappa} H_{n,\kappa}^{\mu\nu}.
\end{equation}

The total angular momentum of a single particle state is $j = \lvert \kappa \rvert - 1/2$ and the orbital angular momentum is given by
\begin{equation}
l = 
\begin{Bmatrix}
\kappa~\mathrm{for}~\kappa > 0 \\
-(\kappa + 1)~\mathrm{for}~\kappa < 0
\end{Bmatrix}
\end{equation}
For the ground state in carbon we assume the lowest energy proton and neutron orbitals are fully occupied. 
These are the $s_{1/2}$ shell $(n=1, \kappa = -1)$ and $p_{3/2}$ shell $(n=1, \kappa = -2)$.

This shell model treatment is known to be a first approximation to the missing-energy distribution.
Experimental data obtained in coincidence experiments, e.g. $(e,e^\prime p)$, show that the discrete states obtain a width, centered around the expected mean-field values~\cite{Dutta:PRCeepAu}. 
This can be implemented empirically by smearing the shell-model states with a Gaussian or Lorentzian~\cite{BENHAR2005, Rocco16, Gonzalez-Jimenez:2021ohu}. 
Additionally, correlations beyond the mean field, both long- and short-range, lead to a partial occupation of the shell model states. This may be taken into account by including spectroscopic factors~\cite{Franco-Munoz:2022jcl, BENHAR2005}. The missing strength then appears at larger missing energies and momenta~\cite{BENHAR:RevModPhyseep} and can be taken into account in factorized approaches, notably in Ref.~\cite{Rocco:2019gfb} for electron-induced pion production. 

\subsection{Nucleon distortion in the RDWIA}
\label{sec:distortion}
The hadron tensor for the interaction with a shell with angular momentum $\kappa$ is
\begin{align}
&H^{\mu\nu}_\kappa = \frac{N_\kappa}{2j+1} \nonumber \\
&\sum_{m_j,s_N} \left[J^\mu\left(m_j,s_N,Q^\mu,k^\mu_N,k^\mu_\pi\right) \right]^\dagger J^\nu\left(m_j,s_N,Q^\mu,k^\mu_N,k^\mu_\pi \right)
\end{align}
where $s_N$ and $m_j$ are the projections of the spin of the final-state nucleon and the angular momentum of the bound state, we average over the $2j+1$ possible states for $m_j$.
The occupation of the state is $N_\kappa$, which within the shell model picture is also $2j+1$.

To make approximations to the hadron current clear, it is instructive to first consider the most general expression for the single-nucleon current in momentum space
\begin{align}
\label{eq:JPspace}
J^\nu &= \frac{1}{\left(2\pi\right)^{3/2}}\int \,d \mathbf{p}^\prime_N \int\,d\mathbf{p}^\prime_\pi \overline{\psi}^{s_N}\left(\mathbf{p}_N^\prime, \mathbf{k}_N \right) \phi^* \left(\mathbf{p}_\pi^{\prime},\mathbf{k}_\pi \right)  \nonumber  \\
&\mathcal{O}^\nu\left(q^\mu, p_N^{\prime}, p^\prime_\pi, p^\prime_m \right) \psi_\kappa^{m_j}\left(\mathbf{p}_m^{\prime} = \mathbf{p}^\prime_N + \mathbf{p}_\pi^\prime - \mathbf{q} \right).
\end{align}
Here $\psi^{s_N}\left(\mathbf{p}^\prime_N, \mathbf{k}_N \right)$ and  $\phi^* \left(\mathbf{p}^\prime_\pi,\mathbf{k}_\pi \right)$ are the outgoing nucleon and pion wavefunctions.
These have fixed asymptotic momenta $\mathbf{k}_N$ and $\mathbf{k}_\pi$ respectively, and are functions of the primed momenta $\mathbf{p}^{\prime}_N$ and $\mathbf{p}^\prime_\pi$.
The bound state wavefunction is $\psi_\kappa$, and the projections of spin and angular momentum of the bound state are denoted by superscripts $s_N$ and $m_j$ respectively. 

The outgoing nucleon and pion are energy eigenstates, their asymptotic momenta $\mathbf{k}$ satisfy the relation $E^2 = \mathbf{k}^2 + M^2$.
In the nuclear interior the particles are not momentum-eigenstates, a momentum operator acting on the wavefunctions yields the primed momenta.
In Eq.~(\ref{eq:JPspace}) the transition operator is hence a function of the primed momenta, these are related by momentum conservation $\mathbf{q} + \mathbf{p}_m^{\prime} = \mathbf{p}^{\prime}_\pi + \mathbf{p}_N^{\prime}$.

The full expression of Eq.~(\ref{eq:JPspace}) is computationally expensive, one has to compute $n_\kappa(2j+1)$ 6-dimensional integrals for every point in the 8 dimensional phase space. Moreover singularities can arise in the pole terms, e.g. in pion exchange contributions, as in general $\mathbf{p}^{\prime 2} \neq \mathbf{k}^2 = E^2 - M^2$. The singularities can be avoided by using as energy of outgoing nucleon and pion in the operator the energy derived from the primed momenta $E \rightarrow E^{\prime 2} = \mathbf{p}^{\prime 2} + M$~\cite{PhysRevC.48.816}.
In this work we make an approximation to the full expression by replacing the primed momenta in the operator (but \emph{only} in the operator) by their asymptotic values
\begin{equation}
\label{eq:OAsympt}
\mathcal{O}^\mu\left(q,p_m^\prime,p_N^\prime,p^{\prime}_\pi \right) \rightarrow \mathcal{O}^\mu\left(q,p_m,k_N,k_\pi \right),
\end{equation}
with $p_m \equiv k_\pi + k_N - q$.
We refer to this as the asymptotic approximation, sometimes called the local approximation~\cite{PhysRevC.48.816}, as it removes derivatives with respect to the coordinates in $r$-space expressions.
We are aware of a limited number of calculations that use the full expression of Eq.~(\ref{eq:JPspace}), these where performed for fully exclusive conditions for knockout from a specific shell in photon-induced reactions~\cite{PhysRevC.48.816, VanderhaeghenPhD}. 
These works seem to imply that the full calculation leads to a slightly more smeared out cross section, in particular for angular distributions, compared to the asymptotic approximation.
We plan to utilize the full calculation, and investigate ambiguities in the transition operator in future works.

With Eq.~(\ref{eq:OAsympt}) one can reduce the expression of Eq.~(\ref{eq:JPspace}) to a single 3-dimensional integral. 
If one writes the momentum-space wavefunctions as the Fourier transform of their coordinate space counterparts one can immediately perform the integrals over the primed momenta, and momentum conservation leads to
\begin{equation}
\label{eq:JAsymptotic}
J^\nu = \int \,d \mathbf{r} e^{i \mathbf{q} \cdot \mathbf{r}} \phi^*(\mathbf{r}, \mathbf{k}_\pi)~\overline{\psi}^{s_N} \left(\mathbf{r},\mathbf{k}_N\right) \mathcal{O}^\nu \psi_\kappa^{m_j}\left(\mathbf{r}\right).
\end{equation}

We will in this work always treat the pion as a plane wave, the final expression for the current in the RDWIA used in this work is then given by
\begin{equation}
\label{eq:JPWpiAsymptotic}
J^\nu = \int \,d \mathbf{r}~e^{i (\mathbf{q} - \mathbf{k}_\pi) \cdot \mathbf{r}}~\overline{\psi}^{s_N} \left(\mathbf{r},\mathbf{k}_N\right) \mathcal{O}^\nu \psi_\kappa^{m_j}\left(\mathbf{r}\right).
\end{equation}
It is clear that Eq.~(\ref{eq:JAsymptotic}) allows to include a distorted pion wavefunction without significant increase of computational cost compared to Eq.~(\ref{eq:JPWpiAsymptotic}).
Instead, the problem is to find a suitable potential to treat the pion wavefunction.
Empirical and microscopic optical potentials derived from fits to pion-nucleus elastic scattering are available, but in these treatments any inelastic rescattering of the pion leads to a loss of flux. Such potentials are suitable to describe the process under exclusive conditions, in which the missing energy of the residual system is restricted to a narrow region.
In neutrino experiments such conditions are not met, instead certain rescattering mechanisms (e.g. absorption) will lead to a reduction of the signal, others (e.g. secondary nucleon knockout) do not, and charge exchange reactions migrate pions from one production channel to another.
As such, an optical potential informed by elastic pion-nucleus scattering would underestimate the total rates in the context of neutrino scattering experiments. 
Contrary to this, the results in which the pion is described by a plane wave in most cases should be expected to overestimate rates in neutrino experiments.

The nucleon states are scattering solutions of the Dirac equation with the real Energy-Dependent RMF (EDRMF) potential introduced in Ref.~\cite{Gonzalez-Jimenez19}.
The EDRMF potential is constructed by scaling the RMF scalar and vector potentials as a function of the nucleon energy, thereby implementing a softening of the potential with increasing energy.
At low energies the potential is identical to the RMF potential used to compute the bound state wavefunctions, thereby the orthogonality of initial and final states is ensured when the momentum content of bound and scattering state could potentially overlap. This ensures specifically that the Pauli principle is satisfied~\cite{Nikolakopoulos19}.
At high energies, cross sections computed with the EDRMF are similar to those obtained with the real part of optical potentials constrained by nucleon-nucleus scattering as shown in Ref.~\cite{Gonzalez-Jimenez:2019ejf}.
We consider the EDRMF potential suitable to describe interactions in which the outgoing nucleon remains undetected (or is not used in the definition of the experimental signal), as is the case in neutrino induced pion production cross sections that we consider. 

To gauge the effect of nucleon distortion we compare the RDWIA calculations with the relativistic plane-wave impulse approximation (RPWIA) where the final state nucleon is described by a plane wave.
In this case the asymptotic evaluation of the operator, Eq.~(\ref{eq:OAsympt}), is of course immediately imposed.
With the plane-wave treatment of the final-state nucleon $\overline{\psi}^{s_{N}} = \overline{u}(\mathbf{k}_N,s_N)e^{-i\mathbf{k}_N\cdot r}$ in Eq.~(\ref{eq:JPWpiAsymptotic}), the integral over $\mathbf{r}$ can be performed immediately resulting in
\begin{equation}
\label{eq:RPWIApspace}
J^\mu = \left(2\pi\right)^{3/2} \overline{u}\left(\mathbf{k}_N,s_N\right)\mathcal{O}^\mu\psi_\kappa^{m_j}(\mathbf{p}_m = \mathbf{k}_\pi + \mathbf{k}_N - \mathbf{q}).
\end{equation}
Here 
\begin{equation}
\psi_\kappa^{m_j}(\mathbf{p}) = \frac{1}{(2\pi)^{3/2}}\int \,d\mathbf{r} e^{-i \mathbf{p} \cdot\mathbf{r}} \psi_\kappa^{m_j}(\mathbf{r})
\end{equation}
is the bound state wavefunction in momentum space.
This expression for the RPWIA matrix elements allows for very efficient calculation of the cross section, and is used to benchmark the robustness of the numerical EDRMF results. 
To do this we perform RPWIA calculations by performing the integral of Eq.~(\ref{eq:JPWpiAsymptotic}) numerically, but with the final-state potentials set to zero. These results differ by less than one percent from the ones obtained with Eq.~(\ref{eq:RPWIApspace}) for all observables presented in this work.

\subsection{Single pion production off the nucleon}
\label{sec:SPPnucleon}
The operator for SPP off the nucleon that is used in the interactions with the nucleus in this work was extensively described in Ref.~\cite{Gonzalez:SPPnucleon}.
The model combines the non-resonant background based on the non-linear sigma model~\cite{Hernandez07}, with the direct and crossed exchange of the $P_{33}(1232)$, $S_{11}(1535)$, $P_{11}(1440)$ and $D_{13}(1520)$ resonances.
The background model at low invariant mass $W \lesssim 1.4~\mathrm{GeV}$ is identical to the model of Hernandez Nieves and Valverde of Refs.~\cite{Hernandez07,Hernandez10,Alvarez-Ruso16}.
At larger invariant masses the model uses the Regge approach described in Ref.~\cite{Gonzalez:SPPnucleon}, in which the tree-level propagator of $t$-channel meson exchanges in the low-energy background are replaced by a Regge propagator.
This approach has been previously applied to electro and photoproduction of pions at high invariant mass~\cite{Kaskulov:Mosel, GVL, Vrancx14a}, and to model the background in electromagnetic meson production analyses~\cite{Aznauryan:2003, Aznauryan:2005, DeCruz12a, Corthals06, Kabirnezhad:2022znc}.
In the region of $1.4~\mathrm{GeV} \lesssim W \lesssim 1.8~\mathrm{GeV}$ a smooth transition between the low-energy and Regge models is implemented~\cite{Gonzalez:SPPnucleon}.

The partially conserved axial current hypothesis (PCAC) is used to determine the axial couplings of the resonances other than the delta from their couplings to the pion, the pseudoscalar form factor is determined by assuming pion-pole dominance. 
The couplings that cannot be determined from PCAC are set to zero, and the $Q^2$ dependence of the axial form factors is assumed to be a modified dipole as in Refs.~\cite{Lalakulich06, Gonzalez:SPPnucleon}.
For the axial coupling of the delta, the model uses the fit obtained by Alvarez-Ruso et al.~\cite{Alvarez-Ruso16}.
This fit introduces a $W$ and $Q^2$ dependent phase in both the vector and axial currents for the Delta such that the total phase of multipole amplitudes can be modified.
The phases are determined such that Watson's theorem~\cite{Watsonstheorem} is satisfied for the specific multipole amplitudes defined in Ref.~\cite{Alvarez-Ruso16}.
We can use this result as our model is practically identical to the HNV model used in this fit at invariant masses below $1.4~\mathrm{GeV}$~\cite{Hernandez:Pion, Hernandez13}.
In Ref.~\cite{Gonzalez:SPPnucleon}, the vector couplings to the resonances determined by Lalakulich et al.~\cite{Lalakulich06} were implemented.
This approach fits experimental data for the helicity amplitudes to parametrize the form factors in the vector current. 
As information on neutron amplitudes was limited, some assumptions on the isoscalar-isovector separation were made in the analysis.
We have revisited the form factors for the higher mass resonances and made small modifications in order to match better with the results for the isovector amplitudes obtained in the MAID07 analysis~\cite{MAID07}.
These updated form factors are described in appendix~\ref{app:helamp}, and a comparison to helicity amplitudes and the MAID07 results is shown.

\begin{figure}
\includegraphics[width=0.49\textwidth]{"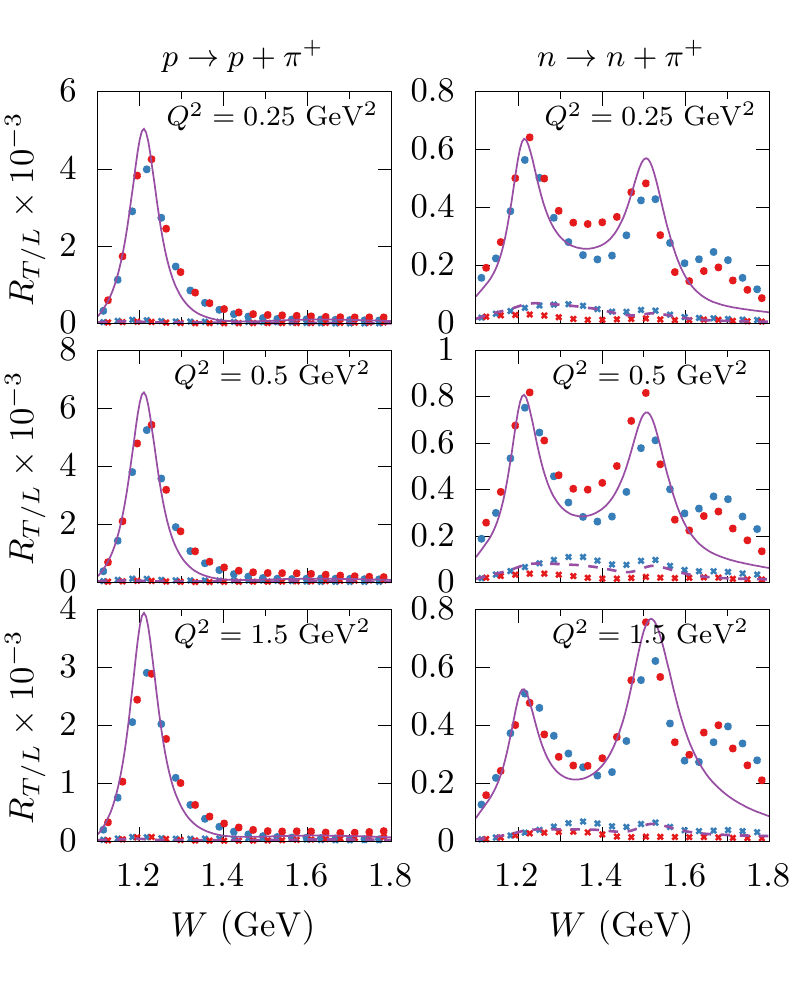"}
\caption{$W$-dependence of the vector-vector contribution to dimensionless responses for different CC interaction channels. The red and blue points are the ANL-Osaka DCC and MAID07 results respectively while the purple lines is the Hybrid model used in this work. The points and solid lines correspond to $R_T^{VV}$ and the crosses and dashed lines represent $R_L^{VV}$.
}
\label{fig:Wdep_CC_Q2}
\end{figure}

We compare the results of the Hybrid model described above to the results of the MAID07~\cite{MAID07} and ANL-Osaka DCC model~\cite{DCC:electron} analyses.
The MAID07 and DCC model results are obtained from the multipole amplitudes that are made available for the different electromagnetic channels~\cite{DCCPWAsite, MAID07site}.
We compute the vector current contribution to the charged current processes by an isospin rotation of the amplitudes for the electromagnetic channels.

In Fig.~\ref{fig:Wdep_CC_Q2} we show the longitudinal and transverse responses integrated over pion angles.
In the limit where the initial and final-state lepton mass can be neglected, and considering the vector current only, the cross section for the charged-current process can be written as
\begin{equation}
\label{eq:Weakresp}
\frac{\diff^2\sigma^{VV}}{\diff W \diff Q^2} = \frac{G_{F}^2\cos\theta_c}{2 E^2\left(2\pi \right)^3}\frac{k_W}{1-\epsilon} \left( R_T^{VV} + \epsilon R_L^{VV} \ \right),
\end{equation}
where $k_W= (W^2-M_N^2)/(2M_N)$.
This expression is similar to the familiar expressions used in electron scattering but with a factor $Q^2$ absorbed in the responses,
\begin{equation}
\label{eq:RLTweak}
R^{VV}_T = \frac{k_\pi^*}{k_W} Q^2 \frac{\left(H^{VV}_{11} + H^{VV}_{22}\right)}{2},~\quad R_L^{VV} = \frac{k_\pi^*}{k_W} Q^2\frac{Q^2}{q^{*2}}H^{VV}_{00}.
\end{equation}
This scaling of the responses better represents the $Q^2$ evolution of the weak cross section, which lacks an overall factor of $(1/Q^2)^2$ compared to electron-induced processes. 
The results for the $W$-dependence of $R^{VV}_{L}$ and $R^{VV}_T$ at different values of $Q^2$ are shown in Fig.~\ref{fig:Wdep_CC_Q2}.
In the cross section for $\pi^+$ production off the proton, which is purely isospin $3/2$ and hence delta-dominated, we find good agreement between the MAID07 and DCC models. The $Q^2$ dependence of the cross section at the delta pole agrees between all models, but the Hybrid model tends to produce a larger peak and smaller tail, also some high-$W$ strength is missing in the model.
In the neutron channel, the discrepancy between the MAID07 and DCC models at higher $W$ is more relevant as the second resonance region contributes with a similar magnitude as the delta for this channel.
The Hybrid model gives respectable results up to the second resonance region, it falls within the difference between MAID07 and DCC model results. 
It resembles the MAID07 model in terms of longitudinal and transverse separation, i.e. a lower $R_T$ with a larger $R_L$ compared to the DCC.

\begin{figure}
\includegraphics[width=0.49\textwidth]{"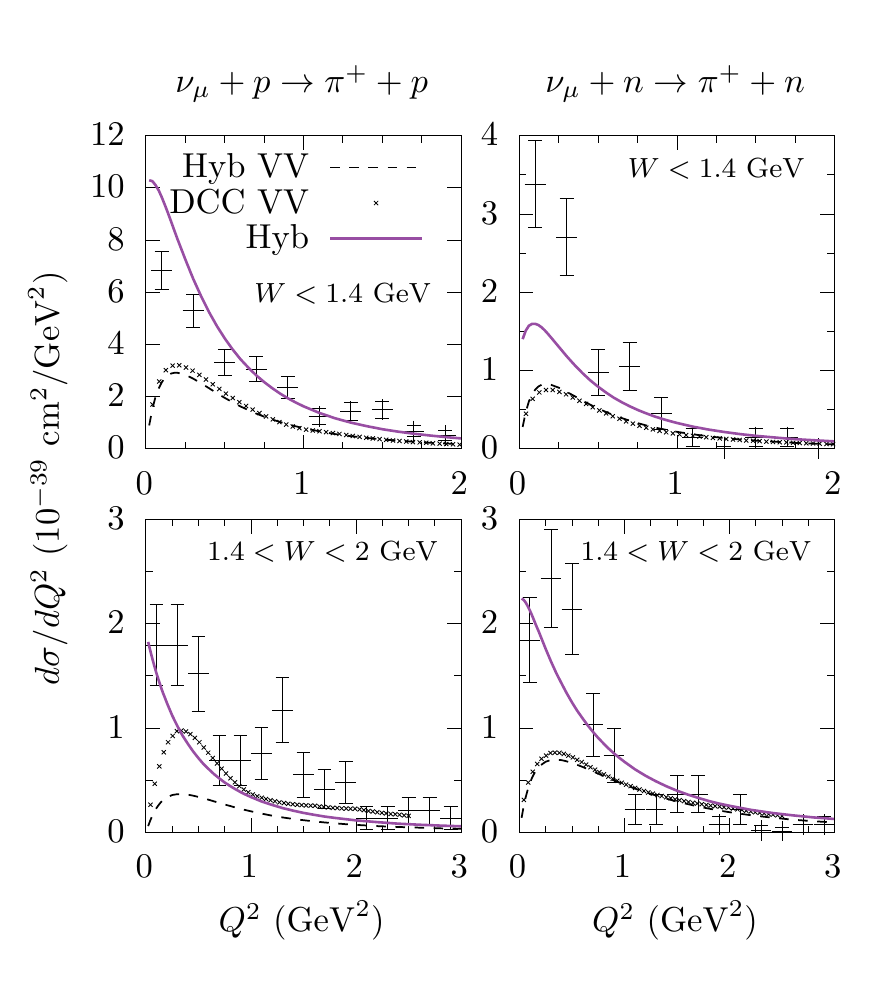"}
\caption{$Q^2$-dependence of the BEBC-flux folded cross sections for charged pion production through CC interactions with the proton and neutron.
The solid purple lines show the total cross section obtained with the Hybrid model, while the dashed line and crosses show the vector-vector contribution to the cross section obtained in the Hybrid model and ANL-Osaka DCC model respectively. The data is from~\cite{ALLASIA1990}}
\label{fig:Qdep_CC}
\end{figure}

\begin{figure}
\includegraphics[width=0.49\textwidth]{"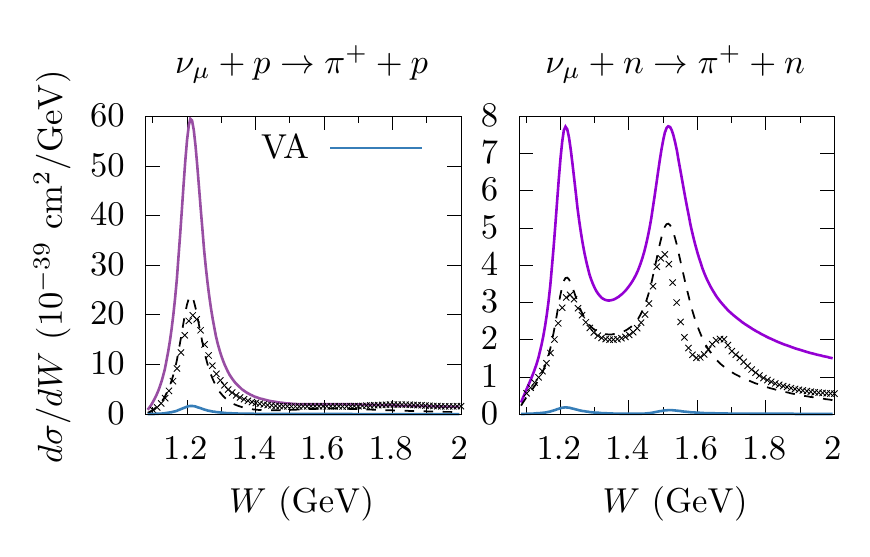"} 
\caption{$W$-dependence of the BEBC-flux folded cross sections for charged pion production through CC interactions with the proton and neutron.
The blue lines show the vector-axial contribution to the cross section computed in the hybrid model, the rest of the lines are the same as in Fig.~\ref{fig:Qdep_CC}.
Both the DCC and Hybrid model results are integrated up to $Q^2 = 2.5~\mathrm{GeV}^2$.}
\label{fig:Wdep_CC}
\end{figure}

It is interesting to consider also cross sections that are flux-folded and partly integrated over lepton kinematics, as one does in a neutrino experiment.
We show calculations for the conditions of the Big European Bubble Chamber (BEBC) experiment in Figs.~\ref{fig:Qdep_CC} and~\ref{fig:Wdep_CC} for the $Q^2$ dependence and the $W$-dependence respectively.
We show separately the vector-vector contribution to the cross section obtained in the Hybrid and DCC models. From Fig.~\ref{fig:Qdep_CC}, one sees that the $Q^2$-dependence is very similar in both models, the only exception is the high-$W$ region for the proton target where the Hybrid model gives smaller results as already seen above.
The results for the $W$-dependence are similar to those shown before, although one sees that the DCC and Hybrid models agree much better in the dip between the Delta and second resonance region when integrated over lepton kinematics. Some compensation between the smaller $R_T$ and larger $R_L$ seems to occur here. One may also notice that the high-$W$ behaviour of the Regge model gives similar results to the DCC.
The BEBC data is particularly interesting because it is taken at incoming energies of several tens of GeV. 
At such high energies the vector-axial interference term becomes negligible (shown by the blue line in Fig.~\ref{fig:Wdep_CC}), and hence the total cross section is composed almost exclusively of the sum of purely vector-vector and purely axial-axial contributions. 
This means that, if one assumes that the DCC model (and thus also the Hybrid model) gives a correct description of the vector contribution, the discrepancy with data in Fig.~\ref{fig:Qdep_CC} can be ascribed to the axial current only.
It is then notable that the vector-vector contribution in the $n\pi^+$ channel seems to be well constrained, while the total cross section is underpredicted by almost a factor $2$ for $W < 1.4~\mathrm{GeV}$ as is also found in the original HNV model~\cite{Hernandez07}.

\section{Results}
\label{sec:results}
We have computed cross sections for $\pi^+$ production of carbon integrated over hadron angles both in the RPWIA and RDWIA as described above.
We perform a direct comparison of both approaches in order to quantify the effect of nucleon distortion for MINER$\nu$A and T2K kinematics.
We first discuss the neutrino flux and kinematic cuts used in obtaining these experimental data.
Then we compare the results to cross sections in terms of lepton kinematics and reconstructed neutrino energies. 
Finally, we pay special attention to the reconstructed $Q^2$ distributions obtained in these experiments.
We discuss nuclear effects which are not included in our analysis, but point out that the nucleon level amplitudes, even in the delta region, should be better constrained before any conclusion can be drawn from the nuclear target data.

\subsection{Measurements by MINER$\nu$A and T2K}
Measurements of neutrino and antineutrino induced charged-pion production on carbon were reported in Refs.~\cite{MINERvACCpi15, MINERvACCpi16, Minerva:CCnuanu2017,MINERvA:pimin2019} by MINERvA and Ref.~\cite{T2K:pipCH} by T2K. 
The interactions are obtained on a predominantly hydrocarbon target which we model as a carbon nucleus and a free proton (CH).
On top of the $\nu$ data we will also include the $\overline{\nu}_\mu$-induced $\pi^-$ production data of Ref.~\cite{MINERvA:pimin2019}, in view of the isospin symmetry between both interactions.
Under the assumption of perfect isospin symmetry for the carbon nucleus, both the neutrino and antineutrino process are described by the same hadron current. The RMF initial state breaks this isospin symmetry, as neutrons are bound more strongly than protons.
The RDWIA includes an additional isospin breaking effect, through the coulomb potential in the final-state.
As these isospin breaking effects are small we will use the current computed for the neutrino-induced $\pi^+$ production on carbon, to describe also anti-neutrino $\pi^-$ production on carbon.

Cross sections for neutrino-induced single $\pi^+$ production in MINER$\nu$A were first reported in Ref.~\cite{MINERvACCpi15},
and an updated dataset was released in Ref~\cite{MINERVACC1pipdata2017}.
We use the updated data, including only the data in which final-states with multiple pions are explicitly rejected.

All datasets include different kinematic cuts, and probe different energy regions.
In the MINER$\nu$A data cuts are performed with \emph{reconstructed} kinematic variables, which depend explicitly on the neutrino energy $E_\nu$.
With knowledge of the incoming energy one can define
\begin{equation}
W^{free} = \sqrt{M_N^2 + 2M_N(E_\nu - E_\mu) - Q^2},
\end{equation}
which is the invariant mass of the hadron system if the interaction occurs on a stationary free nucleon.
The squared four-momentum transfer similarly depends on the incoming energy
\begin{equation}
Q^2 = 2 E_\nu\left(E_\mu - p_\mu\cos\theta_\mu \right) - m_\mu^2.
\end{equation}
It is the reconstructed energy $E_{\nu,(rec)}$, which is inferred from the total visible energy in the detector, from which $Q^2_{(rec)}$ and $W^{free}_{(rec)}$ are computed in the experimental analysis.
Computing these reconstructed variables, while their definition is clear and unambiguous from the experiments point of view, is impossible without repeating a modeling of all visible energy deposits in the detector.
For this reason we treat $Q^2_{rec}$ and $W^{free}_{rec}$ as true variables, meaning that we compute them from the true incoming energy, i.e. we assume that the experimental energy reconstruction is perfect.
All MINER$\nu$A data included here have the restriction $E_{\nu,rec} < 10~\mathrm{GeV}$.
In the $\nu CC1\pi^+$ data $W^{free}_{rec} < 1.4~\mathrm{GeV}$.
For the anti-neutrino data $W^{free}_{rec} < 1.8~\mathrm{GeV}$, and additionally $\theta_\mu < 25~\mathrm{deg}$.
\begin{figure}
\includegraphics[width=0.49\textwidth]{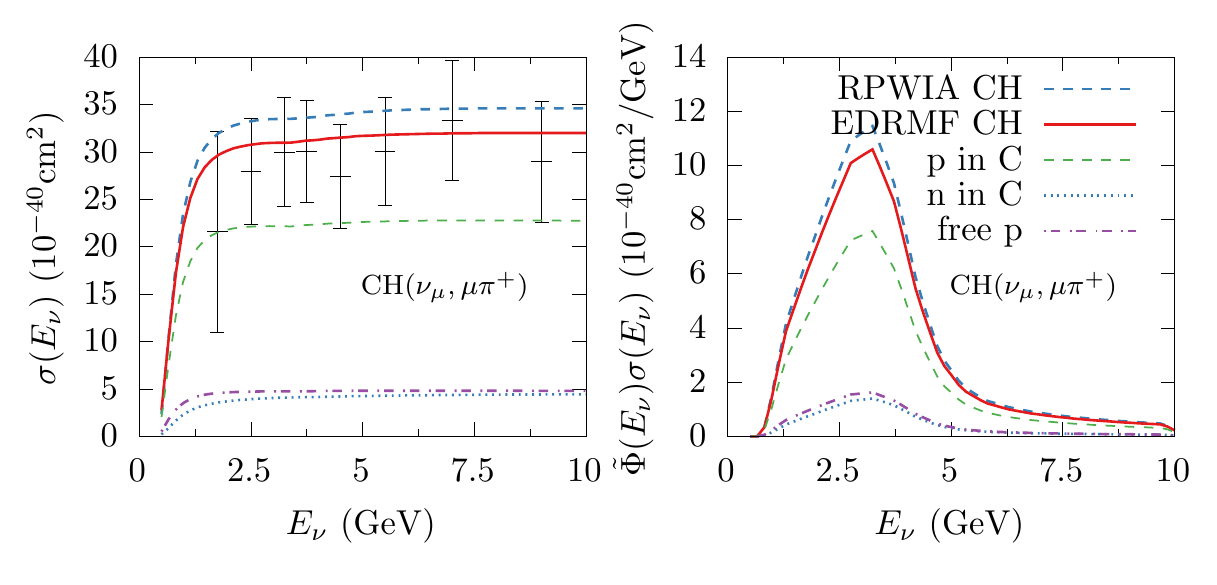} \\
\includegraphics[width=0.49\textwidth]{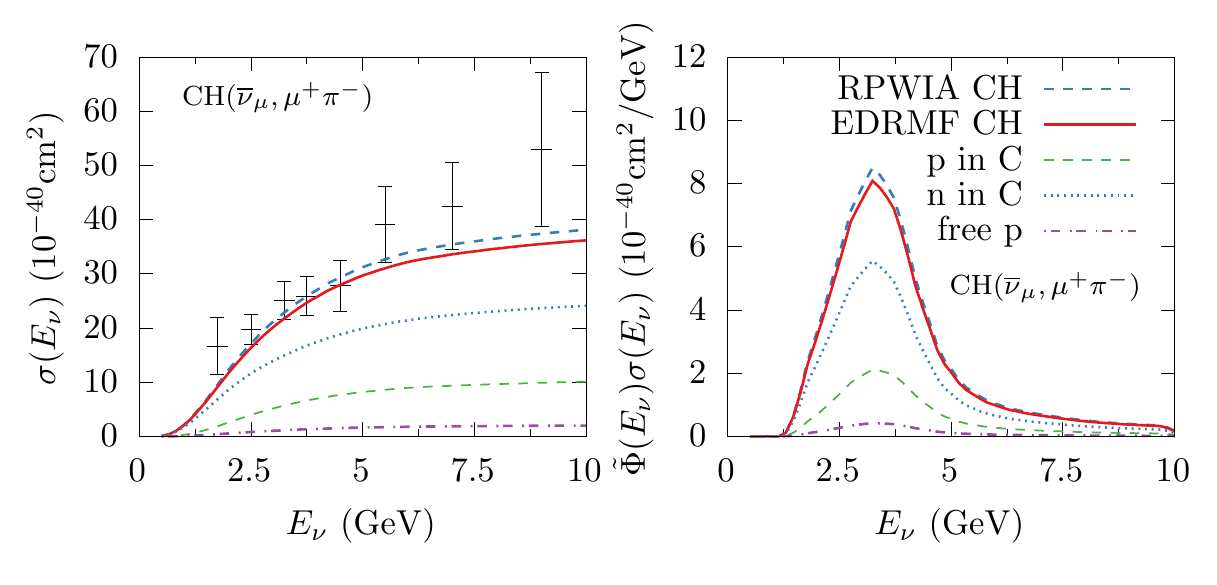} \\
\includegraphics[width=0.49\textwidth]{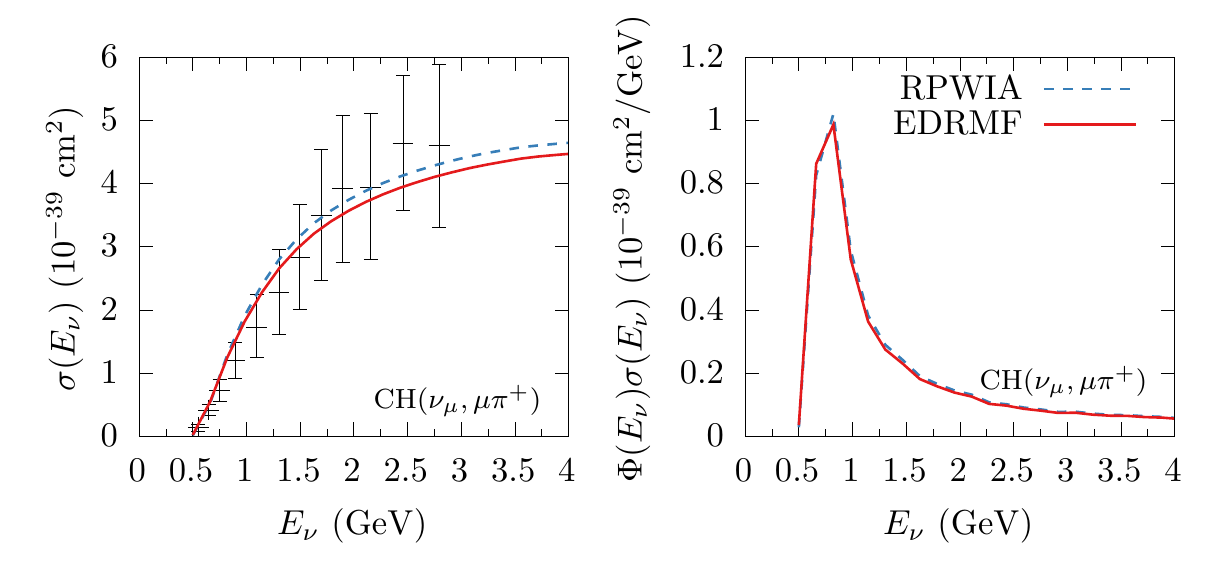}
\caption{Total cross section as function of energy for the MINER$\nu$A CC1$\pi^+$ signal (top), the MINER$\nu$A $\overline{\nu}$CC1$\pi^-$ signal (middle) and the T2K $\nu$CC1$\pi^+$ (bottom). All results are normalized per target nucleon. The corresponding right panels show the cross section weighted with the normalized neutrino flux. The calculations and data for T2K use $\cos\theta_\mu > 0.2$ and $k_\mu > 200~\mathrm{MeV}$.}
\label{fig:Tot_CC1pi}
\end{figure}

The T2K experiment reported a measurement of single $\pi^+$ production on carbon in the T2K near detector in Refs.~\cite{T2K:pipCH, CastilloFernandez:T2KCH}.
In the results reported in Ref.~\cite{T2K:pipCH}, use is made of direct measurements of the pion for which the pion scattering angle with respect to the neutrino beam is restricted to $\cos\theta_\pi > 0.2$.
In Ref.~\cite{CastilloFernandez:T2KCH}, additional distributions are reported which make use of a combination of the direct detection of the pion and the inference of a pion by tagging Michel-electrons from their decays. These are free of cuts on pion angles, and are used in this work.
The T2K measurements only include kinematic cuts on lepton kinematics for some results, which are indicated where appropriate.

\subsection{Cross sections for lepton kinematics}
The results for total cross sections are shown in Fig.~\ref{fig:Tot_CC1pi}. The flux-weighted cross sections are included to show the energy region to which the data is most sensitive.
One observes a reduction of the total cross section in RDWIA compared to the RPWIA. The relative difference between the two is largest for the $\nu CC1\pi^+$ data from MINER$\nu$A, it is seen from the plateau of the cross section that the RPWIA result is approximately 10 percent larger than the EDRMF.
The relative difference in total strength between the RDWIA and RPWIA was found to decrease when a larger region of excitation energy (or equivalently invariant mass) is probed~\cite{Nikolakopoulos:thesis}. Indeed the RDWIA tends to redistribute strength from low to high invariant masses, as such the relative differences are smaller in the $\overline{\nu}CC1\pi^+$ calculations ( $W^{free} < 1.8$) and in the T2K $\nu CC1\pi^+$ results (no invariant mass cuts).

\begin{figure}
\includegraphics[width=0.49\textwidth]{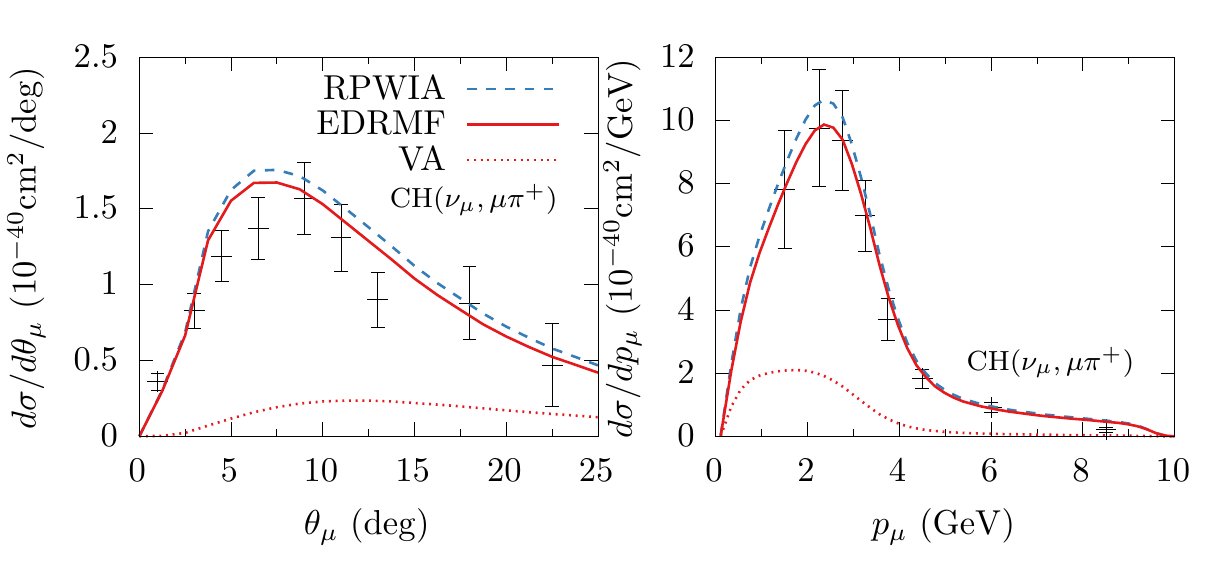} \\
\includegraphics[width=0.49\textwidth]{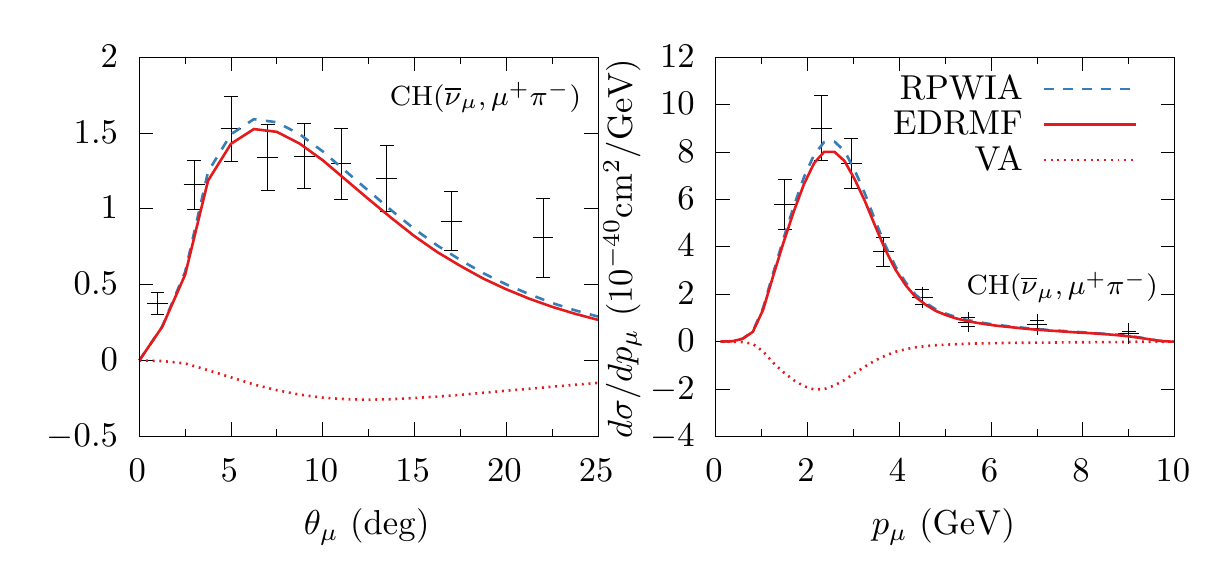}
\caption{ Cross sections differential in lepton kinematics for the MINER$\nu$A CC1$\pi^+$ signal (top panels) and the CC1$\pi^-$ (bottom panels), normalized per nucleon. The vector-axial contribution is shown separately by the dashed lines.}
\label{fig:Lep_Min_CC1pip}
\end{figure}

We show the single differential cross sections in terms of muon momentum and angle for the MINER$\nu$A kinematics in Fig.~\ref{fig:Lep_Min_CC1pip}.
We find an excellent description of the cross sections in terms of muon momentum, which indicates an overall magnitude in line with the data.
We find a fair agreement with the cross section in terms of lepton angles. The data are slightly overpredicted at small angles in the neutrino case, while a slight underprediction of the antineutrino cross section is found for the largest angles.
Note that both the $\nu$ and $\overline{\nu}$ calculations are fully related by isospin symmetry. The reason for the underprediction of the $\overline{\nu}$ is likely that $W_{rec}^{free}$ values up to $1.8~\mathrm{GeV}$ are included, while for neutrinos $W_{rec}^{free} < 1.4~\mathrm{GeV}$. 

\begin{figure}
\includegraphics{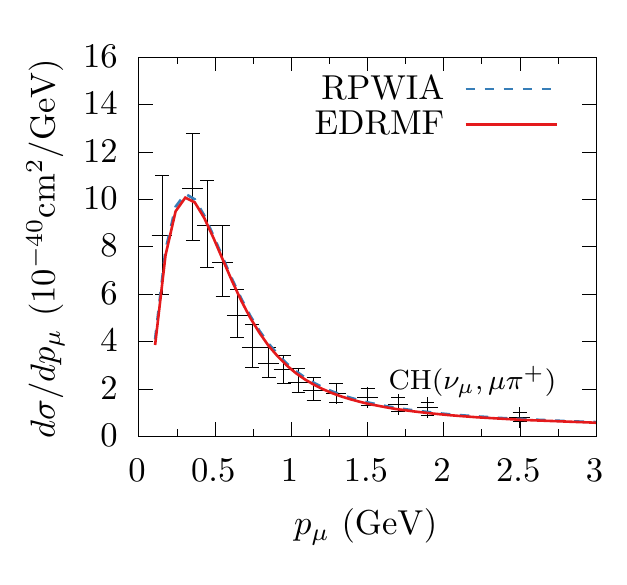}
\caption{Cross section as function of muon momentum compared to the T2K CC1$\pi^+$ data. The data and calculations are for $\cos\theta_\mu > 0.2$. }
\label{fig:Pmu_T2K_pip}
\end{figure}

In Fig.~\ref{fig:Pmu_T2K_pip} we show the results for the momentum of the muon in T2K.
The agreement of the model to the muon data is again very good. The T2K data does not include a cut on the invariant mass, but as the flux peaks at lower energy the experiment is more sensitive to the delta region than MINER$\nu$A is, even without such a cut.

\begin{figure}
\includegraphics[width=0.49\textwidth]{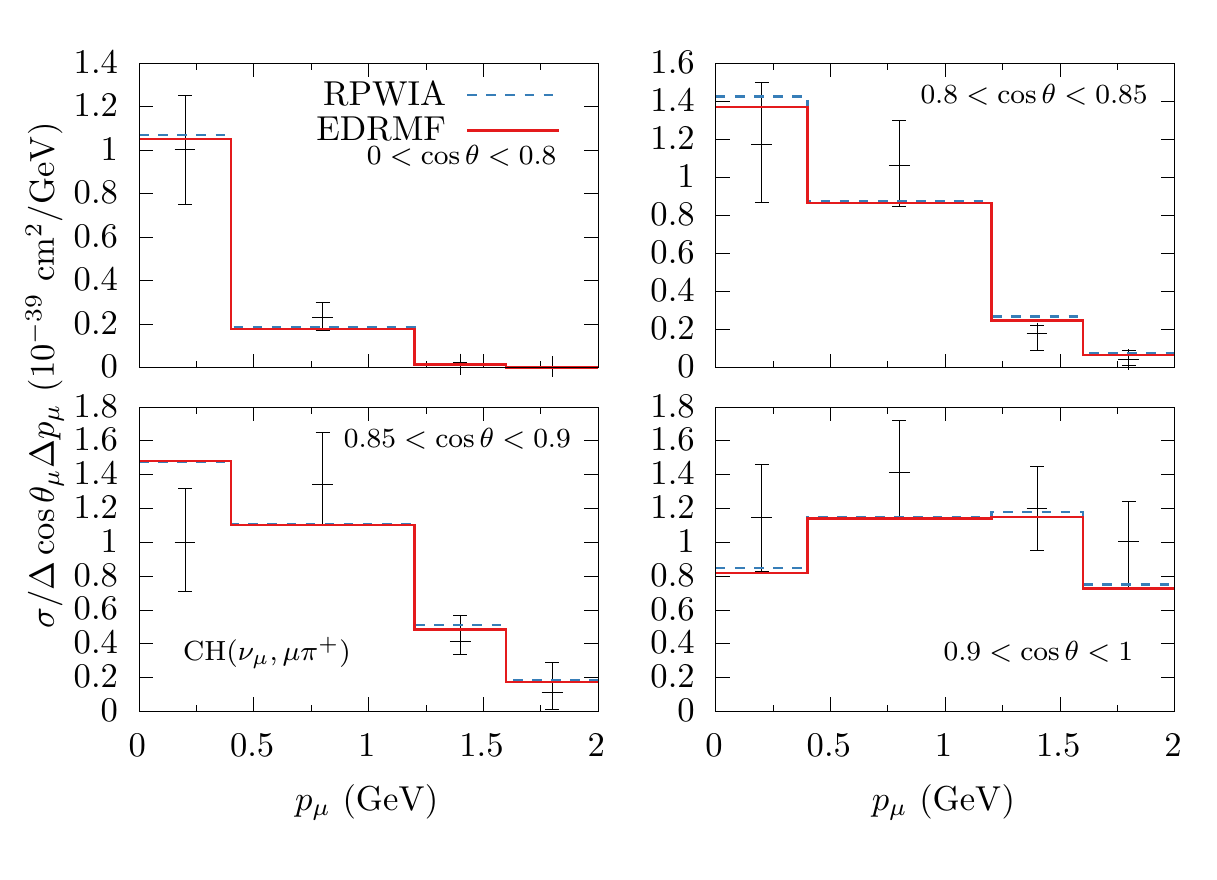}
\caption{Double differential cross section in terms of lepton kinematics compared to the T2K $\nu$CC1$\pi^+$ data, normalized per nucleon.}
\label{fig:DDIF_T2K_CC1pip}
\end{figure}

The T2K collaboration has reported a first measurement of the double-differential cross section in terms of lepton kinematics~\cite{T2K:pipCH}. We show the comparison in Fig.~\ref{fig:DDIF_T2K_CC1pip}. 
The model seems to be in line with the data in most cases although the large error bars allow for significant spread. Even then we find that for some bins the model falls outside of the errorbars. RDWIA and RPWIA results are practically the same, and no clear trend arises in the comparison of the RDWIA and RPWIA in terms of the lepton kinematics.

The EDRMF and RPWIA both yield results for the lepton observables which are in overall agreement with the experimental data. The main exception are the angular distributions, these shape differences are indicative of those found in the $Q^2$ distributions which are discussed in the next section.
One should be prudent in drawing conclusions from the comparisons to data however, as there are a number of caveats which might alter the rates significantly.
On the one hand pion FSI is neglected, the pion wavefunction is treated as a plane wave.
One might expect a further reduction of the cross section when using distorted waves for the outgoing pion, and when inelastic FSI mechanisms such as absorption and charge exchange are taken into account.
The GENIE and NuWro cascade models tend to predict a reduction of the cross sections by $10-20\%$ for MINER$\nu$A kinematics, without significant shape changes for the lepton observables~\cite{MINERvACCpi15, MINERvACCpi16, Minerva:CCnuanu2017,HybridRPWIA}.
Such a decrease could be counteracted by an increase in the neutron-target cross section. Indeed, while the magnitude of the neutrino cross section on the proton is compatible with the data from ANL/BNL and BEBC, the data on neutron targets is underpredicted by almost a factor two even in the $W < 1.4~\mathrm{GeV}$ region, see e.g. results shown in Fig.~\ref{fig:Qdep_CC} and Refs.~\cite{Gonzalez:SPPnucleon, Nikolakopoulos:thesis, Alvarez-Ruso16}.
From Fig.~\ref{fig:Tot_CC1pi}, one sees that the contribution to the total cross section of the neutrons(protons) in neutrino(anti-neutrino) interactions with carbon are significant.

\subsection{$Q^2$ distributions}
\begin{figure*}
\includegraphics[width=0.32\textwidth]{"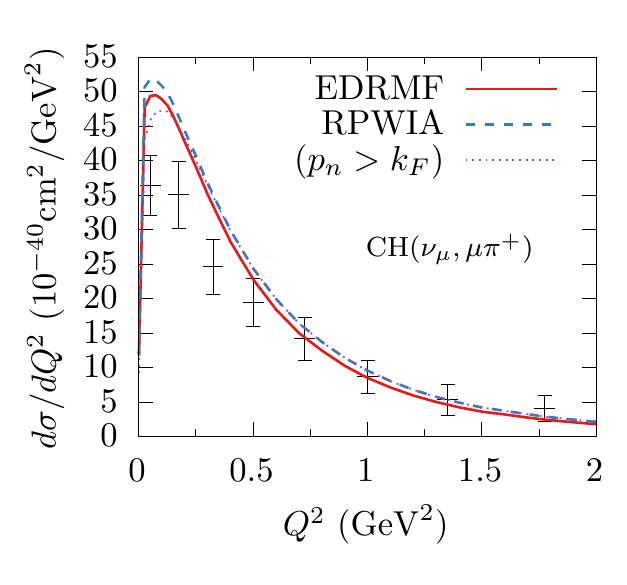"}
\includegraphics[width=0.32\textwidth]{"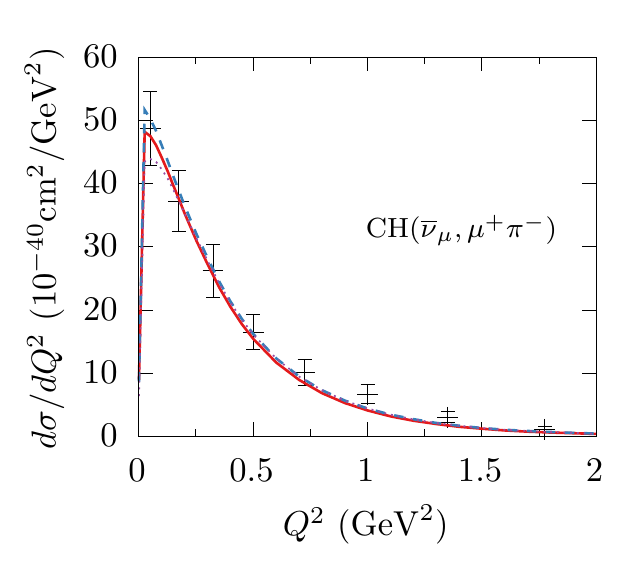"}
\includegraphics[width=0.32\textwidth]{"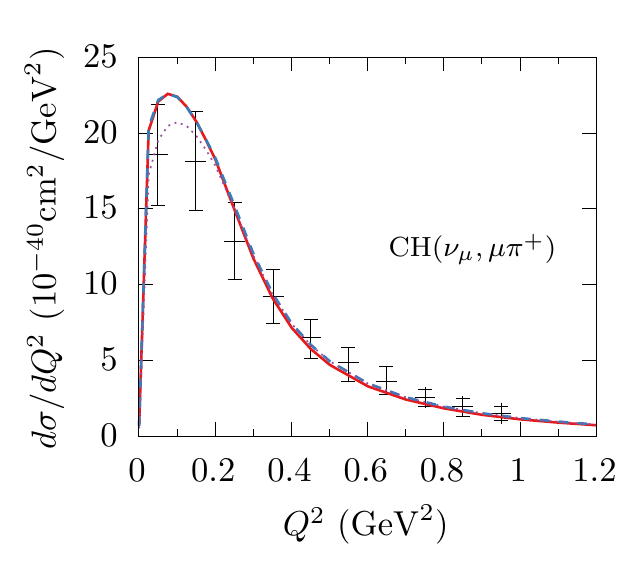"}
\caption{Cross section as function of $Q^2$ for the MINER$\nu$A CC1$\pi^+$ signal on the left,the  MINER$\nu$A CC$1\pi^-$ in the middle and T2K CC1$\pi^+$ the right. The calculation and data for T2K use cuts on lepton kinematics $\cos\theta_\mu > 0.2$ and $p_\mu > 200~\mathrm{MeV}$.}
\label{fig:QQ_FF}
\end{figure*}
Experimental results and analyses of pion production by MINER$\nu$A seem to indicate that data obtained on carbon is in tension with the deuteron target data obtained by ANL/BNL.
In Ref.~\cite{ANLBNLMinerva} a simultaneous fit to the deuteron and different MINER$\nu$A datasets was attempted, and an \emph{ad-hoc} reduction of the cross section at small $Q^2$ for nuclear targets was proposed to resolve tension between the deuteron and MINER$\nu$A results. 
A similar prescription was adopted for the resonant contribution in a fit performed by NO$\nu$A for the inclusive cross section~\cite{NOvA:2020Tune, NOVA:DDIF2021}.
This empirical treatment is motivated by the analogy to quasielastic interactions, where the consistent treatment of initial and final state wavefunctions yields a large reduction of the cross section at low-$Q^2$~\cite{Gonzalez-Jimenez19, Nikolakopoulos19, Jachowicz:JPG2019}. Sometimes this reduction tends to be ascribed solely to collective effects included through the random phase approximation (RPA).
This is somewhat of a misrepresentation as the corrections from RPA yield a much smaller reduction of the cross section when it is implemented with consistent initial and final states~\cite{Nieves:2017lij, Martini:Jachowicz}.
Clearly this empirical treatment is unsatisfactory, and it is important to pin down the source of the discrepancy, be it a nuclear effect or a mismodeling of the hadronic process.

Cross sections as a function of $Q^2$ are shown in Fig.~\ref{fig:QQ_FF} for MINER$\nu$A and T2K datasets.
The comparison of RDWIA and RPWIA results shows that the reduction of the cross section due to nucleon distortion is not specifically confined to the low-$Q^2$ region.
We also show a typical estimation of the effect of Pauli-blocking within the RPWIA.
In these calculations the cross section is set to zero when the outgoing nucleon's momentum is below a fixed Fermi momentum, we use $k_F = 228~\mathrm{MeV}$.
One sees that this procedure results in a suppression only at low-$Q^2$, but it should be considered a crude approximation to implementing the Pauli exclusion principle.
The EDRMF results provide a more realistic treatment of Pauli-blocking as for small values of $T_N$ the initial and final states are orthogonal, for a more detailed discussion see Ref.~\cite{Nikolakopoulos19, Gonzalez-Jimenez19}.
The results for $\pi^+$ production in MINER$\nu$A, which here include the experimental energy spectrum and kinematic cuts, are similar to our previous findings~\cite{Gonzalez-Jimenez19}, where we considered the $Q^2$ dependence at fixed incoming energy.

\begin{figure}
\includegraphics[width=0.48\textwidth]{"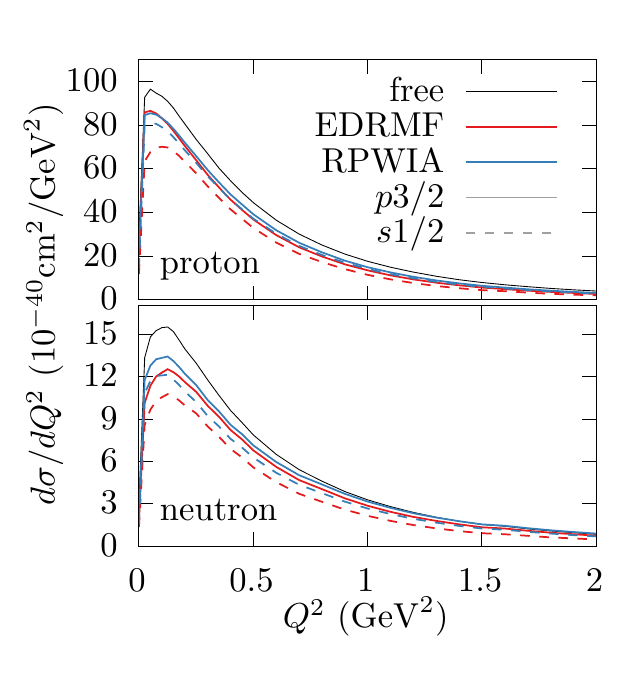"}
\caption{Contribution to the cross section as function of $Q^2$ for the MINER$\nu$A CC1$\pi^+$ signal for each nuclear shell, separately normalized per nucleon. The top panel shows the reaction on protons and the bottom panel on neutrons. We also include the cross section on a free nucleon.}
\label{fig:QQ_shells}
\end{figure}

While we find a small effect of the nucleon distortion for the full experimental signal, we point out that this is not the case for the different contributions to the total cross section.
In Fig.~\ref{fig:QQ_shells}, we show separately the different contributions to the total cross section for the MINER$\nu$A $CC1\pi^+$ signal, normalized per active nucleon.
One sees that the reduction with nucleon distortion compared to the RPWIA is generally larger for the $s\frac{1}{2}$ shell than for the $p\frac{3}{2}$. Additionally the reduction is larger for the interaction on the neutron, than for the proton. Hence the small reduction found in Fig.~\ref{fig:QQ_FF} is a result of the fact that the $p$-shell contribution is double that of the $s$-shell, and that the proton contribution is approximately a factor 6 larger than that of the neutron.

We find an overprediction of the MINER$\nu$A data for $\pi^+$ production at low-$Q^2$, the disagreement is similar in shape to what is found in the GENIE-based analysis~\cite{ANLBNLMinerva}.
The calculation is consistent with the T2K data, for which the signal is dominated by scattering of neutrinos at lower energies.
We cannot conclude that this follows from the energy-weighting however, as the T2K data include different kinematic cuts than the MINER$\nu$A data.
The description of the anti-neutrino data is excellent, although this agreement might deteriorate when FSI are included.

\begin{figure}
\includegraphics[width=0.48\textwidth]{"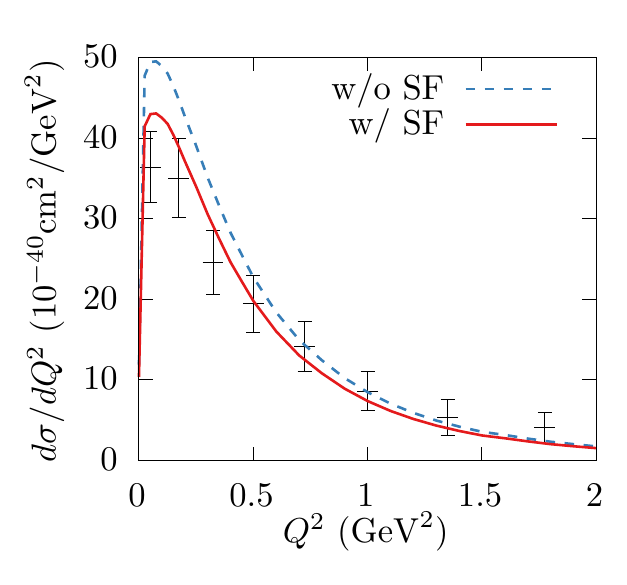"}
\caption{Cross section as function of $Q^2$ for the MINER$\nu$A CC1$\pi^+$ signal, shown with and without spectroscopic factors.
We reduce the occupation of the $p$-shells to $3.3$ nucleons and the occupation of the $s$-shells to $1.8$ nucleons. The result is a flat reduction by a factor $0.87$.}
\label{fig:QQ_SF}
\end{figure}

The treatment of the nucleus is of course not complete, the possibility of modification of resonance properties in the nuclear medium, and the inclusion of a more realistic missing energy-momentum distribution should be considered.
These effects can be included, see e.g.~Refs.~\cite{Gonzalez-Jimenez:2021ohu, Franco-Patino:2022tvv} where a more realistic spectral function is included for one-nucleon knockout, and Refs.~\cite{Nikolakopoulos18a, HybridRPWIA} in which the delta medium modification~\cite{OSMM} is considered within the RPWIA.
An estimate of a calculation with a more realistic spectral function can be done by including a partial occupation of the mean-field states, i.e. spectroscopic factors.
This is shown in Fig.~\ref{fig:QQ_SF}, where we reduce the occupation of the $p\frac{3}{2}$ and $s\frac{1}{2}$ shells to 3.3 nucleons and 1.8 nucleons respectively as in Ref.~\cite{Franco-Munoz:2022jcl}. This leads to a constant reduction of the cross section with a factor $0.87$. This is a lower bound, because the strength missing from the mean-field will contribute at larger missing energies.

Additionally a full treatment of FSI, and notably a realistic treatment of the pion wavefunction is necessary.
If the operator is evaluated at asymptotic momenta, the pion distortion can be included without a significant increase of the computational cost. However, the complicated FSI signatures that should be included make the description of the scattered pion state non-trivial as discussed in section~\ref{sec:distortion}.

Currently available flux-folded neutrino scattering data on nuclei, while obviously important, do not provide the precision required to clearly constrain and separate these effects. 
As such, future work will focus on the description of these effects for the case of pion photo- and electroproduction on nuclei, where more precise constraints can be found.

\begin{figure}
\includegraphics[width=0.49\textwidth]{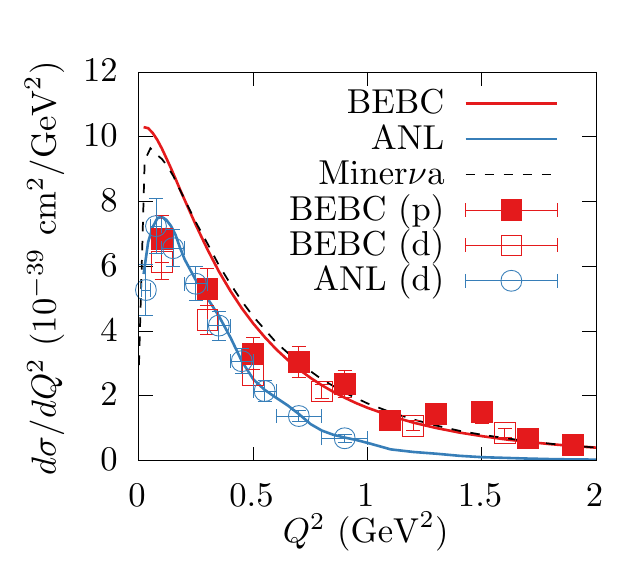}
\caption{Differential cross section as function of $Q^2$ on the proton, obtained for the ANL flux (blue) , the BEBC flux (red), and for the MINER$\nu$A flux and kinematic cuts (dashed black). No deuteron effects are included in these calculations.
The results are compared to the ANL and BEBC datasets on proton and deuteron.
}
\label{fig:Q2BEBCANLpipp}
\end{figure}

\subsubsection{The elephant in the room}
Before ascribing discrepancies to a nuclear effect it is important to consider the accuracy with which the electroweak pion production amplitude on nucleons is known.
A major uncertainty comes from the axial couplings to the resonances. 
In the MINER$\nu$A $\pi^+$ case, with the cut $W^{free} < 1.4~\mathrm{GeV}$, the delta is the most relevant.
The couplings to the delta used in this work, and in many other studies, are determined from the ANL/BNL data for pion production on the deuteron~\cite{CC-ANL82, CC-BNL86}.
The bubble chamber data are limited by statistics however, and significant uncertainties related to the treatment of the deuteron FSI~\cite{Nakamura:deuteronANL}, and the knowledge of the absolute flux~\cite{Wilkinson14} are not fully under control~\cite{NuSTEC:2020nsl, NuSTECWP}.
Because of the limitations of the datasets, models for the axial couplings to the delta have to make some assumptions.
In the following we briefly discuss said assumptions, and describe the procedure followed in Ref.~\cite{Alvarez-Ruso16}, to fit the delta coupling to the ANL data.

Within the isobar model the $N$-$\Delta$ excitation through the axial current is parametrized by 4 form factors, assumed to be functions of $Q^2$ only.
The vertex function, equivalent to the one defined for the vector current in appendix~\ref{app:helamp}, is 
\begin{align}
\Gamma^{\alpha\mu}_A &= \frac{C_3^A}{\mu} \left(g^{\alpha\mu}\slashed{q} - q^\alpha \gamma^\mu \right) + \frac{C_4^A}{\mu^2}\left( g^{\alpha\mu} q\cdot k_R - q^\alpha k_R^\mu \right) \nonumber \\
 &+C_5^A g^{\alpha\mu} + \frac{C_6^A}{\mu^2} q^\alpha q^\mu,
\end{align}
where the scale to make the form-factors dimensionless is set $\mu = M_N$ for the following discussion.
PCAC and pion pole dominance motivate the relation
\begin{equation}
C_6^A(Q^2) = C^A_5(Q^2) \frac{M_N^2}{M_\pi^2 + Q^2},
\end{equation}
for the pseudoscalar form factor.
Quark model results when the delta is considered as an $s$-wave state and dynamical calculations, both imply that $C_3^A = 0$~\cite{Hemmert:1995, ADLER1968, Sato:2021pco}.
This leaves $C_5^A$ and $C_4^A$ to be determined.
In the present results the constraint 
\begin{equation}
C_4^A(Q^2) = -C_5^A(Q^2)/4,
\end{equation}
derived from Ref.~\cite{ADLER1968}, is imposed~\cite{Hernandez07, Alvarez-Ruso16}.
The quark model results of Ref.~\cite{Hemmert:1995}, see also discussion in Ref.~\cite{Sato:2021pco}, imply a similar proportionality between these terms
\begin{equation}
C_4^A = - \frac{M_N^2}{M_\Delta\left(M_N + M_\Delta\right)} C^A_5 \approx -C^A_5/3.
\end{equation}
The proportionality between $C_4^A$ and $C_5^A$ together with $C_3^A = 0$ can be considered an equivalent in the axial current to magnetic multipole dominance in the vector current, both follow from the $s$-wave quark model~\cite{Hemmert:1995}.

The $C_5^A(Q^2$) coupling with the above constraints was fit to the ANL dataset in Ref.~\cite{Alvarez-Ruso16}. In this fit Watson's theorem was imposed for the dominant vector and axial-vector multipoles, and deuteron effects were taken into account within the PWIA~\cite{Watsonstheorem, Hernandez10}.
The resulting model, with a dipole for $C_5^A(Q^2)$, yields a value for $C_5^A(0)$ consistent with the Goldberger-Treiman relation, and provides a good description of the $\pi^+$ production cross section of ANL with an invariant mass cut of $W < 1.4~\mathrm{GeV}$.

However, the same parameters fail to describe the BEBC data on hydrogen at low $Q^2$.
We illustrate this in Fig.~\ref{fig:Q2BEBCANLpipp} through the comparison of calculations for the ANL and BEBC fluxes.
One sees that while the model agrees with the ANL data, the BEBC dataset is overpredicted at low-$Q^2$.
The discrepancy is similar in shape to what is found in comparison to the MINER$\nu$A data. We have added in Fig.~\ref{fig:Q2BEBCANLpipp} the cross section off hydrogen computed with the MINER$\nu$A flux, the similarity to the BEBC-flux averaged result is evident.

One can easily estimate the magnitude of terms involving $C_3$ and $C_4$ from e.g. the expressions provided in Ref.~\cite{Lalakulich06}.
One finds that their contributions are small compared to the $(C_5)^2$ term, and that variations on the assumptions mentioned above cannot readily explain this discrepancy.

To our knowledge, there is no good reason to discredit the BEBC dataset. Or rather, no reason which would not also apply to the ANL dataset. 
It is clear that unless this discrepancy in the proton/deuteron data is resolved one cannot ascribe the low-$Q^2$ discrepancies found in MINER$\nu$A and NO$\nu$A solely to a nuclear effect.
Modern neutrino experiments with proton and deuteron targets could help to resolve uncertainties that plague the current datasets.
Such experiments would prove invaluable to pin down the delta coupling, the far less constrained couplings to higher mass resonances, and the axial coupling to the nucleon~\cite{Alvarez-Ruso:2022ctb}.

\section{Conclusions}
\label{sec:conclusions}
We have performed calculations for neutrino and anti-neutrino induced charged pion production on carbon for the large phase space spanned by the MINER$\nu$A and T2K experiments in the relativistic distorted wave impulse approximation (RDWIA). 
We included the distortion of the outgoing nucleon wavefunction by treating it as a solution of the Dirac equation in the energy-dependent relativistic mean field potential of Ref.~\cite{Gonzalez-Jimenez19}.
This approach ensures consistency of initial and final-states at small nucleon energy, and hence naturally incorporates the Pauli-exclusion principle. 
The real potential includes the necessary of final-state interactions when the outgoing nucleon remains undetected, and may or may not rescatter.
Indeed, this approach provides an excellent description of inclusive $(e,e')$ as seen in Refs.~\cite{Gonzalez-Jimenez19, Gonzalez-Jimenez:2019ejf}.

For the single-nucleon operator of electroweak pion production we use the model of Ref.~\cite{Gonzalez:SPPnucleon}, which is an extension of the HNV model of Refs.~\cite{Hernandez:Pion, Alvarez-Ruso16, Hernandez:PionNucleus,Sobczyk:angles} to higher invariant masses by including additional resonances and a Regge approach for the background contribution at high-$W$. 
In previous works~\cite{Gonzalez:SPPnucleon, Nikolakopoulos18a, HybridRPWIA} we used the vector form factors for the higher mass resonances determined in Ref.~\cite{Lalakulich06}.
We have made modifications to the form factors in this work, in particular to improve the description of the isovector form factors, for which we use results from Refs.~\cite{MAID07, Hernandez08}.
We benchmark the vector-vector contribution to the cross section through comparison with the MAID07~\cite{MAID07} and ANL-Osaka dynamic coupled channels~\cite{DCC:electron} models.
Although the model we use is simple, and the description of the vector current is not complete, we find reasonable agreement with the MAID07 and DCC models up to the second resonance region.
This is in particular the case for inclusive cross sections at high energies, where the vector-vector contribution to the cross section agrees to within a couple of percent with the ANL-Osaka result for $W < 1.4~\mathrm{GeV}$.

We compare the RDWIA results with the relativistic plane-wave impulse approximation (RPWIA) in which the distortion of the final state nucleon is neglected.
The RDWIA leads to a reduction of the total cross section of up to $10 \%$ compared to the RPWIA, but no significant change in shape is found for flux-averaged observables measured in T2K and MINER$\nu$A.

We find that both the RDWIA and RPWIA results are consistent with the T2K and MINER$\nu$A data, apart from the $\pi^+$ production cross section at low-$Q^2$ measured in MINER$\nu$A, the latter is overpredicted by both approaches.
An overprediction of similar shape and size of $\pi^+$ production data is also found in the MINER$\nu$A and NO$\nu$A analyses which both use some variant of the GENIE event generator to model the cross section.
Both in Ref.~\cite{ANLBNLMinerva}, and in the fits performed by NO$\nu$A~\cite{NOvA:2020Tune}, an \emph{ad-hoc} suppression of the cross section at low $Q^2$ is introduced, which is ascribed to an unspecified nuclear effect.
The treatment is motivated by analogy to quasielastic interactions where indeed the consistent treatment of initial and final state wavefunctions, and further collective effects included through e.g. the RPA, lead to a reduction of the cross section at low-$Q^2$.
In this work we have included a consistent treatment of nucleon states within the RDWIA, and do not find a reduction of the cross section specifically at low-$Q^2$.

We show that an overprediction of the cross section of similar shape and size at low-$Q^2$ is also present in comparison to the BEBC data on a hydrogen target.
Unless the discrepancy between the results for ANL and BEBC kinematics is resolved, or the BEBC data can be rightfully ignored, one cannot ascribe the discrepancies found in the MINER$\nu$A and NO$\nu$A solely to a nuclear effect.
One should give similar weight to the idea that the axial couplings to the nucleon are not sufficiently constrained by the ANL data alone.
Constraints on the axial form factors might come from ChPT~\cite{Yao:2018pzc, HillerBlin:2016hau, GuerreroNavarro:2019fqb}, quark-hadron duality~\cite{Sato:2021pco, SajjadAthar:2020nvy}, or progress in lattice QCD~\cite{Barca:2021iak, Ruso:2022qes}. 
Theoretical advances should ideally be supported by new measurements of neutrino scattering on proton and deuteron targets.
Such experiments would prove invaluable to pin down the delta coupling, the far less constrained couplings to higher mass resonances, and the axial coupling to the nucleon~\cite{Alvarez-Ruso:2022ctb}.

 The description of the nuclear matrix elements is not complete, we discuss the prospect of further considering the effect of pion FSI, correlations beyond the mean field, possible medium modifications of resonances, and the asymptotic approximation of the single-nucleon operator used in this work.
Current neutrino-nucleus datasets are not suitable for validation of nuclear models, as the neutrinos span a broad energy range, and the underlying couplings to the nucleon are not well known.
Instead future efforts will focus on the description of electron and photoproduction datasets, for which the single-nucleon operator can be better constrained.

\acknowledgments 
This paper was authored by the Fermi Research Alliance, LLC under Contract No. DE-AC02-07CH11359 with the U.S. Department of Energy, Office of Science, Office of High Energy Physics.
This research was funded by the Research Foundation Flanders (FWO-Flanders), the government of Madrid and Complutense university under project PR65/19-22430 (R.~G.-J).
The computational resources (Stevin Supercomputer Infrastructure) and services used in this work were provided by Ghent University, the Hercules Foundation, and the Flemish Government.

\onecolumngrid

\newpage
\appendix
\section{Vector Form factors}
\label{app:helamp}
We list the parametrizations of the vector-current form factors for the $S_{11}$, $P_{11}$, and $D_{13}$ used in this work. All other form factors are the same as in Ref.~\cite{Gonzalez:SPPnucleon}.
We compare the helicity amplitudes obtained with these form factors to those of the MAID07 analysis~\cite{MAID07}, and the results of CLAS analyses compiled in Ref.~\cite{MokeevHAmp}.

The helicity amplitudes are computed in the reference system where a resonance with mass $M_R$ is produced at rest.
The relevant four-vectors are explicitly given by
\begin{equation}
k_{\gamma^{*}} = q =
\begin{pmatrix}
\omega \\
0 \\
0 \\
\lvert\vec{q}\rvert
\end{pmatrix}, 
k_{N} =
\begin{pmatrix}
M_R - \omega \\
0 \\
0 \\
-\lvert\vec{q}\rvert
\end{pmatrix}, 
k_{R} =
\begin{pmatrix}
M_R \\
0 \\
0 \\
0
\end{pmatrix},
\end{equation}
for the (virtual) photon, the nucleon, and the produced resonance respectively. The helicity amplitudes are defined as
\begin{align}
\label{eq:A12}
\mathcal{A}_{1/2} = \sqrt{\frac{2\pi\alpha}{K}}\frac{1}{e} \langle S_{z,R} = \frac{1}{2} \rvert \epsilon_\mu^{(+)}(q) J^\mu \lvert S_{z,N} = -\frac{1}{2} \rangle, \\
\label{eq:A32}
\mathcal{A}_{3/2} = \sqrt{\frac{2\pi\alpha}{K}}\frac{1}{e} \langle S_{z,R} = \frac{3}{2} \rvert \epsilon_\mu^{(+)}(q) J^\mu \lvert S_{z,N} = \frac{1}{2} \rangle, \\
\label{eq:S12}
\mathcal{S}_{1/2} = \sqrt{\frac{2\pi\alpha}{K}}\frac{1}{e} \langle S_{z,R} = \frac{1}{2} \rvert \frac{\lvert\vec{q}\rvert}{\sqrt{Q^2}}\epsilon_\mu^{(0)}(q) J^\mu \lvert S_{z,N} = \frac{1}{2} \rangle,
\end{align}
where as usual $Q^2 = -q^2$, $\epsilon_\mu^{(\pm)} = \mp \frac{1}{\sqrt{2}}\left(0, -1, \mp i, 0 \right)$, and $\epsilon_\mu^{(0)} = \frac{1}{\sqrt{Q^2}}\left(\lvert\vec{q}\rvert, 0 , 0, -\omega \right)$.

\subsection{Spin-$1/2$ Resonances}
We use the following parametrization of the current
\begin{equation}
\label{eq:curr12}
\frac{1}{e}\langle S_{z,R} \rvert J^\mu \rvert S_z \rangle = \overline{u}(k_{R},S_{z,R})~\left[\frac{F_1}{\mu^2}\left(q^\mu\slashed{q} - q^2\gamma^\mu\right)  + i\frac{F_2}{\mu} \sigma^{\mu\alpha}q_{\alpha}\right] \begin{Bmatrix} \gamma^5 \\ 1 \end{Bmatrix} u(k_N, S_z),
\end{equation}
with the $\gamma^5$ for abnormal parity transitions $1/2^+ \rightarrow 1/2^-$ (the $S_{11}$), and the unit matrix for normal parity transitions $1/2^+ \rightarrow 1/2^+$ (the $P_{11}$).
We use $\mu = M_N + M_R$ for the form factors presented below.
The Dirac spinors are 
\begin{equation}
u(\mathbf{k}, S_z) = \sqrt{\frac{E+M}{2M}}
\begin{pmatrix}
\phi(S_z) \\ 
\frac{\mathbf{\sigma}\cdot\mathbf{k}}{E+M} \phi(S_z),
\end{pmatrix}
\end{equation}
where
\begin{equation}
\phi(S_z = +1/2)=
\begin{pmatrix}
1 \\
0
\end{pmatrix},
~\phi(S_z = -1/2)=
\begin{pmatrix}
0 \\
1
\end{pmatrix}.
\end{equation}
By using the Gordon identity to rewrite the $i2\sigma^{\mu\nu} = -\left[\gamma^\mu, \gamma^\nu \right]$ term, Eq.~(\ref{eq:curr12}) can be written as
\begin{equation}
\frac{1}{e}\langle S_{z,R} \rvert J^\mu \rvert S_z \rangle = \overline{u}(k_{R},S_{z,R})~\left[\frac{F_1}{\mu^2}\left(q^\mu\slashed{q} - q^2\gamma^\mu\right)  + \frac{F_2}{\mu} \left( \gamma^\mu(M_R \mp M_N) - k_N^\mu - k_{R}^\mu \right) \right]\begin{Bmatrix} \gamma^5 \\ 1 \end{Bmatrix} u(k_N, S_z),
\end{equation}
which can be compared to the expressions given in Ref.~\cite{Aznauryan:2008}. One has the following relation at $W=M_R$ between the set of form factors used here and the form factors $G_1,~G_2$ defined in Refs.~\cite{Devenish:1975jd, Aznauryan:2008}
\begin{equation}
\label{eq:F_vs_G_12}
 \frac{F_1}{\mu^2} = \mp G_1,~\mathrm{and}~\frac{F_2}{\mu} = \pm\frac{M_R \pm M_N}{2}G_2.
\end{equation}
With the parametrization of Eq.~(\ref{eq:curr12}) the helicity amplitudes are 
\begin{align}
\label{eq:A_12_FF}
\mathcal{A}_{1/2}  &= \sqrt{\frac{4\pi\alpha}{K}} \sqrt{\frac{E_N\pm M_N}{2M_N}} \left[ \frac{F_1}{\mu^2} Q^2 + \frac{F_2}{\mu}\left(M_R \mp M_N \right) \right], \\
\mathcal{S}_{1/2}  &= \mp \sqrt{\frac{2\pi\alpha}{K}} \sqrt{\frac{E_N \pm M_N}{2M_N}} \lvert \vec{q} \rvert \left[ \frac{F_1}{\mu^2} \left( M_R \mp M_N \right) -\frac{F_2}{\mu} \right].
\label{eq:S_12_FF}
\end{align}
These relations are the same as those given in Refs.~\cite{Hernandez08, Leitner09}. 
Compared to Lalakulich et al.~\cite{Lalakulich06}, our expression for the transverse amplitude $\mathcal{A}_{1/2}$ is the same, while the scalar amplitude $\mathcal{S}_{1/2}$ differs by a factor $\frac{\lvert \vec{q}^{CMS} \rvert}{\lvert \vec{q}^{LAB} \rvert} = \frac{M_N}{M_R}$.
This is the case because the amplitudes of Ref.~\cite{Lalakulich06} are computed in the lab-frame, as pointed out previously in Ref.~\cite{Leitner09}.
We present amplitudes determined in the resonance rest frame.
This means that when the same form factors as in Ref.~\cite{Lalakulich06} are used, the scalar amplitudes will be smaller than in the original publication.
\begin{figure*}
\includegraphics[width=0.48\textwidth]{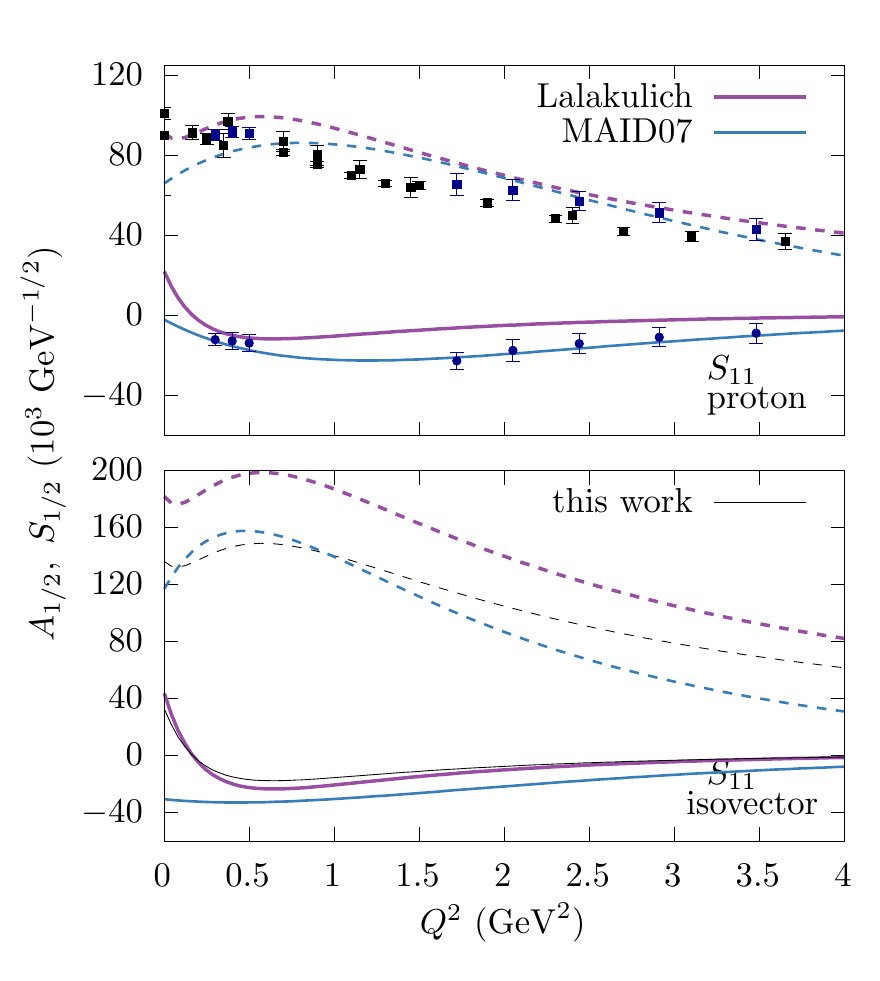} 
\includegraphics[width=0.48\textwidth]{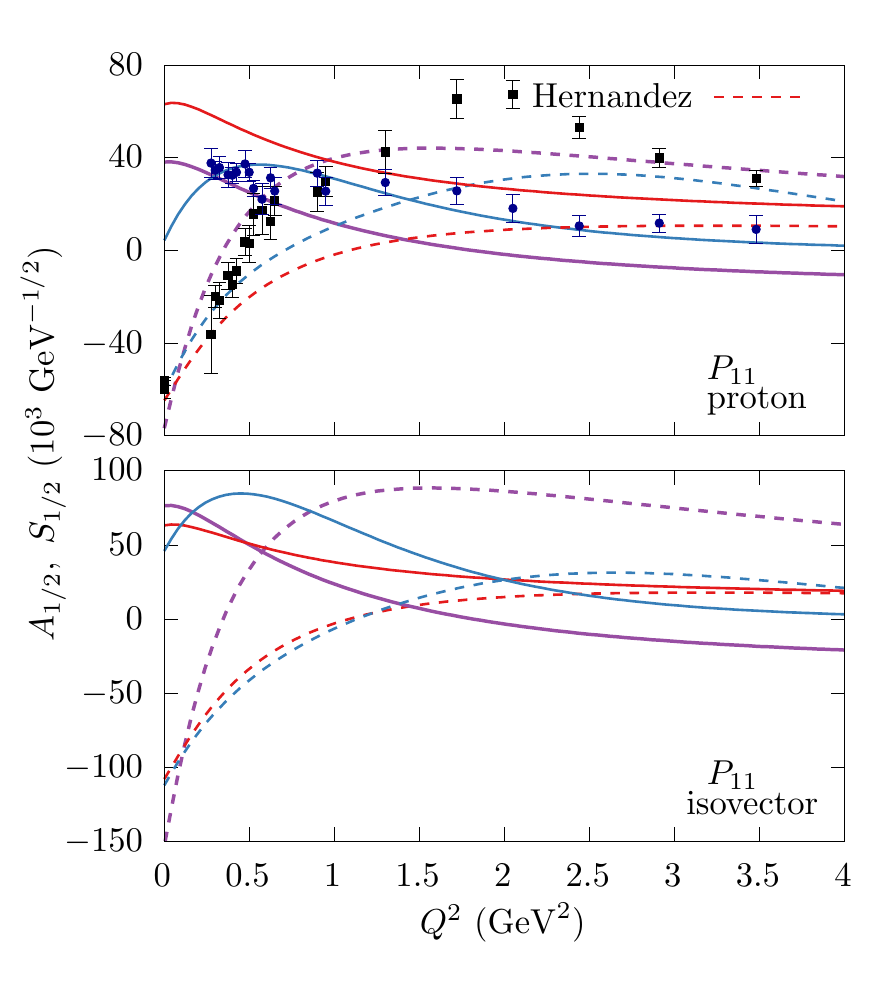}
\caption{Helicity amplitudes $A_{1/2}$ (dashed lines, squares) and $S_{1/2}$ (solid lines, circles), obtained with the form-factors of Lalakulich~\cite{Lalakulich06}, from the MAID07 analysis~\cite{MAID07}, and from the form factors of Hernandez et al.~\cite{Hernandez08}.
The top panels show the proton amplitudes and the bottom panels the isovector amplitudes $A_{p} - A_{n}$. The points are the results of CLAS analyses of Refs.~\cite{CLAS:photopion2009, PhysRevC.86.035203, CLAS:epi2009, CLAS:e2pi2016, CLAS.S11.eta:2007, CLAS.S11.eta:2001}. 
For the $S_{11}$ only one analysis~\cite{CLAS:epi2009}, shown by the blue points, separates scalar and vector amplitudes. The black points are the results~\cite{CLAS.S11.eta:2007, CLAS.S11.eta:2001} where $S_{1/2}=0$ is assumed.}
\label{fig:Spin12_HA}
\end{figure*}

The helicity amplitudes obtained with several models are compared to the results of analyses compiled in Ref.~\cite{MokeevHAmp} in Fig.~\ref{fig:Spin12_HA}.
We use the parametrization of Lalakulich for the form factors for $S_{11}$ production off the proton in this work, these are given by
\begin{align}
F_1^p(Q^2) &= \frac{2G_D(Q^2)}{1+Q^2/(1.2M_V^2)} \left[1 + 7.2\ln\left(1 + Q^2/(\mathrm{GeV}^2) \right) \right], \\
F_2^p(Q^2) &= 0.84 G_D(Q^2) \left[1 + 0.11\ln\left(1 + Q^2/(\mathrm{GeV}^2) \right) \right],
\end{align}
where $G_D(Q^2) = \left(1 + Q^2/M_V^2 \right)^{-2}$ and $M_V = 840~\mathrm{MeV}$.
With these form-factors one finds reasonable agreement with the proton-target amplitudes from the MAID07 analysis.
Both parametrizations tend to overestimate the results of the CLAS analyses. It is notable that the latter includes data for $\eta$ production~\cite{CLAS.S11.eta:2007, CLAS.S11.eta:2001}, where additionally the assumption $\mathcal{S}_{1/2}=0$ is made.
This additional data reduces the magnitude of transverse amplitude compared to data originally used in the fit of Ref.~\cite{Lalakulich06}.
In previous work~\cite{Gonzalez:SPPnucleon, Nikolakopoulos18a}, we used the assumption of a negligible isoscalar contribution, meaning that $F^n = -F^p$. 
The resulting isovector amplitudes are compared to the MAID07 result in Fig.~\ref{fig:Spin12_HA}.
The MAID07 results imply that the neutron amplitudes are smaller and drop off more rapidly with $Q^2$ than the proton amplitudes.
In this work we use the simple relation 
\begin{equation}
F^n = -1/2 F^p
\end{equation}
for both $F_1$ and $F_2$.
This brings $\mathcal{A}_{1/2}$ closer to the MAID07 analysis in the $Q^2$ region of interest.
It should be noted that MAID07 uses different values of the width, $\pi N$ branching ratio of the $S_{11}$, and a significant phase at resonance position. 
If we absorb the ratios of these parameters in the form factors, the results for the inclusive cross sections shown in Fig.~\ref{fig:Wdep_CC_Q2} are significantly larger than both the MAID07 and ANL-Osaka results, hence we do not rescale the form factors as such. 

For the $P_{11}$ we use the parametrization of Hernandez et al.~\cite{Hernandez08}, which is given by
\begin{align}
F_{1}^p(Q^2) &= \frac{-5.7 G_D(Q^2)}{1+Q^2/1.4M_V^2}, \\
F_2^p(Q^2) &= -0.64G_D(Q^2)\left(1 - 2.47\ln \left( 1 +\frac{Q^2}{\mathrm{GeV}^2}  \right) \right),
\end{align}
for the proton form factors.
The assumptions made in Ref.~\cite{Hernandez08} for the neutron amplitudes is that $\mathcal{A}^n_{1/2} = -2/3A^p_{1/2}$ and $S^n_{1/2}=0$.
For the isovector form factor this results in
\begin{align}
F_{1}^V &= \frac{F_1^p \left( (M_N+M_R)^2 + 5Q^2/3 \right) + 2/3 F_2^p \left( M_N + M_R \right)\mu }{ (M_N+M_R)^2 + Q^2}, \\
F_{2}^V &= \frac{F_2^p \left( 5(M_N+M_R)^2 - 3Q^2 \right)\mu - 2 F_1^p Q^2 \left( M_N + M_R \right)}{3\mu\left( (M_N+M_R)^2 + Q^2 \right)},
\end{align}
where we fix $M_R = 1.44~\mathrm{GeV}$ for all kinematics.

The helicity amplitudes for the $P_{11}$ are shown the right panels of Fig.~\ref{fig:Spin12_HA}.
We note that in Ref.~\cite{Hernandez08} a minus sign is included in the scalar amplitude with respect to Eq.~(\ref{eq:S_12_FF}) when comparing to certain datasets (e.g. MAID07).
We include this minus sign in the scalar amplitude shown in Fig.~\ref{fig:Spin12_HA}.
The $\mathcal{S}_{1/2}$ obtained with the Lalakulich form factors changes sign around $Q^2 = 1.5~\mathrm{GeV}^2$ as do the helicity amplitudes to which this fit was performed.
In the more recent compilation shown here~\cite{CLAS:epi2009}, the scalar amplitude at high $Q^2$ has similar magnitude but the opposite sign.
The assumption $F^p = -F^n$ is used in the Lalakulich form factors.
This leads to a large $\mathcal{A}_{1/2}$, compared to the other parametrizations.

\subsection{Spin-$3/2$}
The vector current contribution of spin-$3/2$ resonances is parametrized as 
\begin{equation}
\label{eq:curr32}
\frac{1}{e} \langle S_z^* \rvert J^\mu \lvert S_z \rangle = \overline{\psi}_\alpha \left(k_R, S_{z,R} \right) \Gamma^{\alpha\mu}_V \begin{Bmatrix} \gamma^5 \\ 1 \end{Bmatrix} u\left(k_N, S_z\right),
\end{equation}
where the top corresponds to abnormal parity transition $1/2^+ \rightarrow 3/2^+$ (the delta resonance) and the bottom to normal parity transitions $1/2^+ \rightarrow 3/2^-$ (the $D_{13}$ resonance).
The vertex factor is 
\begin{equation}
 \Gamma_V^{\alpha\mu} = \left[ \frac{C_3}{\mu} \left(g^{\alpha\mu} \slashed{q} - q^\alpha\gamma^\mu\right) +
 \frac{C_4}{\mu^2} \left(g^{\alpha\mu} q\cdot k_{R} - q^\alpha k_{R}^\mu\right) + 
 \frac{C_5}{\mu^2} \left(g^{\alpha\mu} q\cdot k_{N} - q^\alpha k_{N}^\mu\right) \right],
\end{equation}
where $\mu$ is an arbitrary scale to make the form factors dimensionless.
Comparing this to the definition of $G_1, G_2, G_3$ of Devenish~\cite{Devenish:1975jd} one finds following relations between the two sets of form factors
\begin{equation}
G_1 = - \frac{C_3}{\mu}, ~ G_2 = \mp \frac{(C_4 + C_5)}{\mu^2}, ~ G_3 = \pm \frac{C_5}{\mu^2}.
\end{equation}
The Rarita-Schwinger spinors are defined as $\psi^\mu(k,S) = \sum_{\lambda,s} \cgc{\frac{1}{2}}{1}{s}{\lambda}{\frac{3}{2}}{S} u\left(k,s\right) \epsilon^\mu \left(k, \lambda \right)$ with the brackets denoting the Clebsch-Gordan coefficients. 
Where the polarization vectors for the resonance in its rest frame are
\begin{align*}
\left(\epsilon^0_\mu\right)^* &= \frac{1}{M_{R}} \left(\lvert\vec{k_{R}}\rvert, 0, 0 , -E_{R} \right) = \left(0,0,0,-1\right), \nonumber \\
\left(\epsilon^{(\pm)}_\mu\right)^* &= \mp \frac{1}{\sqrt{2}} \left(0, -1, \pm i, 0\right).
\end{align*}
From these expressions, and with the definition of the current in the (ab)normal parity case of Eq.~(\ref{eq:curr32}) one finds following results for the helicity amplitudes 
\begin{align}
\label{eq:A32_FF_32}
\mathcal{A}_{3/2} &= \pm  \sqrt{\frac{2\pi\alpha}{K}\frac{(E^*_N \mp M_N)}{2M_N}} \left[\frac{C_3}{\mu} \left(M_{R} \pm M_N \right) + \frac{C_4}{\mu^2} q\cdot k_{R} + \frac{C_5}{\mu^2} q\cdot k_N \right], \\
\label{eq:A12_FF_32}
\mathcal{A}_{1/2} &= \sqrt{\frac{1}{3}} \sqrt{\frac{2\pi\alpha}{K}\frac{(E^*_N \mp M_N)}{2M_N}} \left[\frac{C_3}{\mu}\frac{\left(Q^2 + M_N(M_N \pm M_R) \right)}{M_R} - \frac{C_4}{\mu^2}q\cdot k_R - \frac{C_5}{\mu^2} q\cdot k_N \right],  \\
\label{eq:S12_FF_32}
\mathcal{S}_{1/2} &= \mp \frac{\lvert \vec{q}^* \rvert}{M_R} \sqrt{\frac{2}{3}} \sqrt{\frac{2\pi\alpha}{K}\frac{(E^*_N \mp M_N)}{2M_N}} \left[\frac{C_3}{\mu}M_R + \frac{C_4}{\mu^2}{M_R^2} + \frac{C_5}{\mu^2}M_R\left(M_R - \omega^*\right) \right].
\end{align}
Again, there is a difference of a factor $M_N / M_R$ compared to Ref.~\cite{Lalakulich06} as explained above for the spin-$1/2$ case. 

For the Delta resonance we retain the parametrization used in Refs.~\cite{Lalakulich06,Hernandez:Pion,Gonzalez:SPPnucleon}.
This is consistent with the fit of the axial coupling to the delta and delta phases of Ref.~\cite{Alvarez-Ruso16} that are used in this work.

\begin{figure*}
\includegraphics[width=0.48\textwidth]{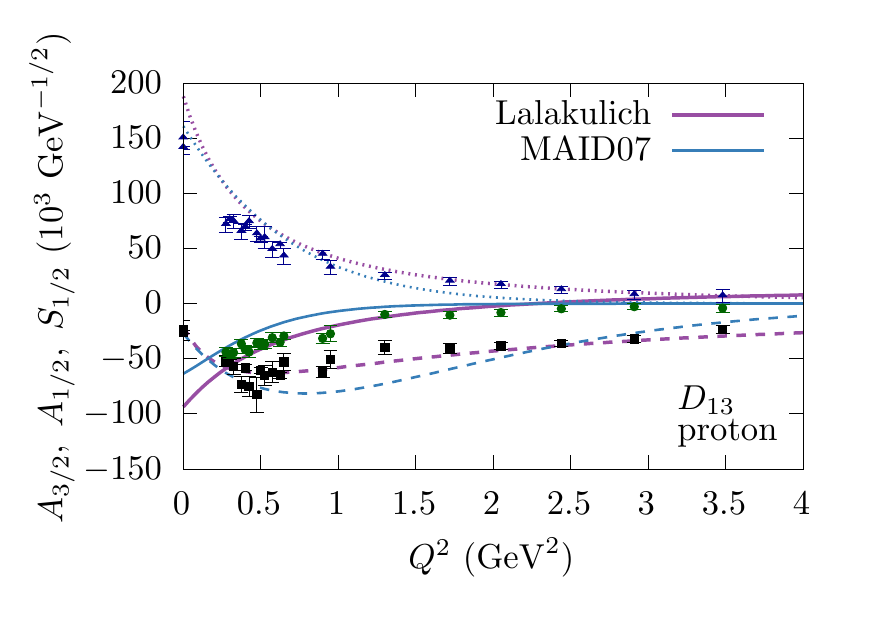} 
\includegraphics[width=0.48\textwidth]{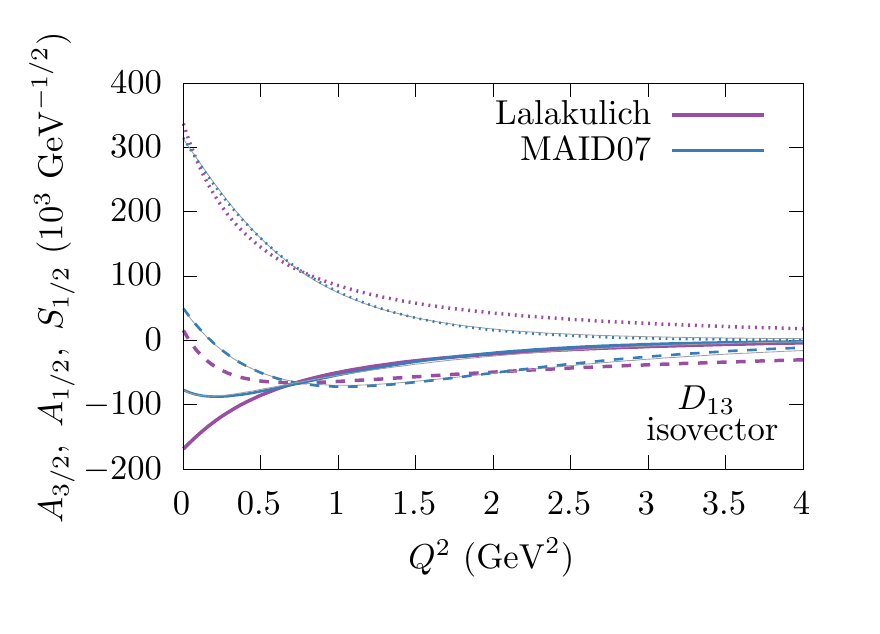}
\caption{Helicity amplitudes $A_{3/2}$ (dotted lines, triangles), $A_{1/2}$ (dashed lines, squares), and $S_{1/2}$ (solid lines, circles). 
The left panel shows the proton amplitudes and the right panel the isovector amplitudes $A_{p} - A_{n}$. The points are the results of analyses of CLAS data~\cite{CLAS:photopion2009, PhysRevC.86.035203, CLAS:epi2009, CLAS:e2pi2016}.
The dark-gray lines in the right panel are the fit to the MAID07 results of Eqs.~(\ref{eq:C3Vfit}-\ref{eq:C5Vfit}).}
\label{fig:Spin32_HA}
\end{figure*}

Amplitudes for the $D_{13}$ resonance are shown in Fig.~\ref{fig:Spin32_HA}.
We note that several works include a minus sign in the definition of $\mathcal{S}_{1/2}$ compared to Eq.~(\ref{eq:S12_FF_32}), this is in particular the case for the MAID07 result, and the analysis of Ref.~\cite{CLAS:epi2009}.
We follow Refs.~\cite{Leitner09, Hernandez:PionNucleus} and take this change of sign into account. 
To do this we invert the system of Eqs.~(\ref{eq:A32_FF_32}-\ref{eq:S12_FF_32}), including a minus sign in the scalar amplitude, to obtain numerically the form factors implied by the MAID07 amplitudes.
We then fit these form factors as follows, with $\mu = M_N$,
\begin{align}
\label{eq:C3fit}
C_3^p &= -2.72~G_{D}\left(Q^2,1.53\right) e^{-0.38 Q^2/\mathrm{GeV}^2}, \\
\label{eq:C4fit}
C_4^p &= 3.13~G_{D}\left(Q^2,0.84\right)  e^{-0.66 Q^2/\mathrm{GeV}^2},  \\
\label{eq:C5fit}
C_5^p &= -1.66~G_{D}\left(Q^2,0.80\right) e^{-0.96 Q^2/\mathrm{GeV}^2}\left(1-2.513 Q^2/\mathrm{GeV}^2 \right),  \\
\label{eq:C3Vfit}
C_3^V &= -3~G_{D}\left(Q^2,2.00\right)    e^{-0.54 Q^2/\mathrm{GeV}^2}, \\
\label{eq:C4Vfit}
C_4^V &= 4.73~G_{D}\left(Q^2,1.13\right)  e^{-0.73 Q^2/\mathrm{GeV}^2},  \\
\label{eq:C5Vfit}
C_5^V &= -3.65~G_{D}\left(Q^2,0.99\right) e^{-0.97 Q^2/\mathrm{GeV}^2}\left(1-1.150 Q^2/\mathrm{GeV}^2 \right).
\end{align}
Here $G_D(Q^2,x) = \left(1+Q^2/(xM_V^2) \right)^{-2}$ is a modified dipole and the superscript $V$ denotes the isovector form factor.
We use these form factors for the $D_{13}$ throughout this work. They are similar to the parametrization of Ref.~\cite{Hernandez:PionNucleus}, which was used in the fit of the delta coupling of Ref.~\cite{Alvarez-Ruso16}. The helicity amplitudes that result from this fit are shown by gray lines in Fig.~\ref{fig:Spin32_HA}, they are practically indistinguishable from the MAID07 result.

\bibliographystyle{apsrev4-1.bst}
\bibliography{bibliography}

\end{document}